\documentclass[reprint, NumberedRefs]{JASA}

\usepackage{hyperref}
\hypersetup{colorlinks,breaklinks,
             urlcolor=blue, linkcolor=blue, citecolor = blue}

\usepackage{float}

\newcommand{\tc}[1]{\textcolor{blue}{#1}}

\usepackage{booktabs}

\begin{document}

\title[RTF Simulation]{Simulating room transfer functions between transducers mounted on audio devices using a modified image source method}

\author{Zeyu Xu}
\email{zeyu.xu@audiolabs-erlangen.de}
\thanks{ORCID: 0000-0002-4158-6218}
\affiliation{International Audio Laboratories, Am Wolfsmantel 33, 91058, Erlangen, Germany.}
\thanks{A joint institution of the Friedrich-Alexander-Universit\"{a}t Erlangen-N\"{u}rnberg (FAU) and Fraunhofer IIS, Germany.}
\author{Adrian Herzog}
\affiliation{International Audio Laboratories, Am Wolfsmantel 33, 91058, Erlangen, Germany.}
\author{Alexander Lodermeyer}
\affiliation{Fraunhofer Institute for Integrated Circuits IIS, Am Wolfsmantel 33, 91058, Erlangen, Germany.}
\author{Emanu\"{e}l~A.\,P.~Habets}
\affiliation{International Audio Laboratories, Am Wolfsmantel 33, 91058, Erlangen, Germany.}
\author{Albert G. Prinn}
\affiliation{Fraunhofer Institute for Integrated Circuits IIS, Am Wolfsmantel 33, 91058, Erlangen, Germany.}

\date{\today} 


\begin{abstract} 
 
The image source method (ISM) is often used to simulate room acoustics due to its ease of use and computational efficiency. The standard ISM is limited to simulations of room impulse responses between point sources and omnidirectional receivers. In this work, the ISM is extended using spherical harmonic directivity coefficients to include acoustic diffraction effects due to source and receiver transducers mounted on physical devices, which are typically encountered in practical situations. The proposed method is verified using finite element simulations of various loudspeaker and microphone configurations in a rectangular room. It is shown that the accuracy of the proposed method is related to the sizes, shapes, number, and positions of the devices inside a room. A simplified version of the proposed method, which can significantly reduce computational effort, is also presented. The proposed method and its simplified version can simulate room transfer functions more accurately than currently available image source methods and can aid the development and evaluation of speech and acoustic signal processing algorithms, including speech enhancement, acoustic scene analysis, and acoustic parameter estimation. 

\end{abstract}


\maketitle

\section{Introduction}

Room acoustic simulation has been an active research field since the 1960s~\cite{Schroeder1961}. It is of paramount importance for, e.g., auralization and virtual acoustics~\cite{kleiner1993, Savioja_1999, Schroder2011, Vorlander2013, Schissler2016, Vorlaender2020, Tang2021}, evaluation of acoustic signal processing algorithms~\cite{Mabande2013, Cherkassky2017, Evers2018, Das2021, Herzog2022a}, and data generation for data-driven methods~\cite{Diaz2020, Gelderblom2021, Huebner_et_al_2021, Wechsler2022, Srivastava2022, Srivastava_et_al_2023, Herzog2022b, Grumiaux2022}. Room acoustics simulation methods can be broadly categorized into geometric methods~\cite{Allen1979, Peterson1986, Torres2018, Tang2020a, bezzam2020study}, wave-based methods~\cite{Murphy2000, Kowalczyk2010, Sakuma_et_al_2014, Hamilton_and_Bilbao_2017, Hargreaves_et_al_2019, Yoshida_et_al_2022}, hybrid methods~\cite{Aretz2012, Thomas2017}, and, more recently, deep-learning-based methods~\cite{Tang2021, Ratnarajah2021b, Luo2022learning}. This paper presents an extension to the well-known geometrical method, the image source method (ISM), which enables the inclusion of acoustic diffraction effects from audio devices of finite extent in shoebox-shaped rooms. 

Modern smart speakers are a typical embodiment of an audio device that contains loudspeakers and microphones. Generating a reverberant room acoustic simulation environment which includes the directivities of loudspeakers and microphones due to the transducer properties and the acoustic diffraction effects, is essential for developing and testing new acoustic signal processing algorithms. When using wave-based methods, the acoustic diffraction effects caused by the loudspeaker enclosure are implicitly modeled. However, because of the much higher computational cost of solving wave-based models, geometrical methods are often preferred. While assuming that sound waves behave like light rays does not apply to long wavelengths, Aretz \textit{et al.} \cite{aretz2014application} have shown that the ISM can be used to approximate wave-based method solutions above the Schroeder frequency~\cite{Schroeder1962} in rectangular rooms by specifying angle-dependent reflection coefficients. Thus, the ISM can be used to obtain acceptable solutions across most of the audible frequency range above the Schroeder frequency. 

The ISM has been implemented in many widely used toolboxes~\cite{Habets2006, Pyroomacoustics2018, Wabnitz2010, gpuRIR2020}, along with various approaches for reducing the computational complexity, for example, parallelization based on GPUs \cite{fu2016gpu,gpuRIR2020}. Directional properties of the source and receiver were also incorporated into the ISM to produce more realistic room impulse responses~\cite{Kompis1993, Habets2006, Wabnitz2010, Betlehem2012, Pyroomacoustics2018, Brinkmann2018}, and efforts have been made to include acoustic diffraction effects due to the human head \cite{Kompis1993} and rigid spherical baffles \cite{Jarrett2012b, Hafezi2015}. Jarrett \textit{et al.}~\cite{Jarrett2012b} extended the ISM to include rigid spherical microphones using an analytical expression of the diffraction effects in the spherical harmonic domain. Samarasinghe \textit{et al.}~\cite{samarasinghe2018spherical} further extended the ISM, viz. the Generalized ISM (GISM), to facilitate parameterization of the reverberant transfer function between a directional source and a directional receiver. More recently, an extension of the ISM to incorporate loudspeaker directivities in the far field in the spherical harmonic domain has been reported~\cite{luo2021fast}. However, the existing approaches cannot simulate a transfer function between a directive source and a directive receiver that includes near-field diffraction effects. 

In this work, the GISM~\cite{samarasinghe2018spherical} is extended to enable the incorporation of acoustic diffraction effects. First, we review the basic concepts of the ISM and its extension to the spherical harmonic domain in Sec.~\ref{sec:2}. In Sec.~\ref{sec:3}, we present our proposed room transfer function (RTF) parameterization, which utilizes the source directivity coefficients computed from solutions of an exterior radiation problem and receiver directivity coefficients obtained using reciprocity~\cite{Xu_et_al_2022}. In addition, we derive a simplified version of the proposed method to reduce computational complexity. The Finite Element Method (FEM) is used to verify the proposed method by comparing RTFs with the source and receiver mounted on either different devices or on the same device. The experimental setups and the evaluation of the proposed method are presented in Sec.~\ref{sec:4}. 
The effect of varying the parameters of the device, e.g., its position in the room, physical shape, and size, on the accuracy of the proposed method is investigated. Finally, conclusions are drawn in Sec.~\ref{sec:5}. 

\section{The Image Source Method}
\label{sec:2}

\subsection{Standard image source method}
\label{sec:21}

\begin{table*}[t]
	\caption{Parameters and descriptions used in ISM} 
	\begin{tabular}{l l}
		\toprule
		Parameter & Descriptions  \\	
		\midrule
		$L_x, L_y, L_z$ & Room dimensions: length, width and height \\ 
		$\beta_{x1},\beta_{x2},\beta_{y1},\beta_{y2},\beta_{z1},\beta_{z2}$ & Reflection coefficients of the walls \\
		$\mathbf{p} = (p_x,p_y,p_z) \in \mathcal{P}$ & Triple determining whether the image source is mirrored \\ 
		~ & w.r.t. walls at $x=0$, $y=0$ or $z=0$, elements take values of $0$ or $1$  \\
		$\mathbf{q} = (q_x,q_y,q_z) \in \mathcal{Q} $ & Triple determining higher order reflections \\
		~ & each element can take values between $[-N_m, N_m]$ \\ 
		$\mathbf{x}^{\text{sI}}_{\mathbf{p},\mathbf{q}} $ &  Source images \\
		$\mathbf{x}^{\text{rI}}_{\mathbf{p},\mathbf{q}} $ &  Receiver images \\ 
		$\mathbf{R}^{\text{sI}\to\text{r}}_{\mathbf{p},\mathbf{q}}=\mathbf{x}_\text{r}- \mathbf{x}^{\text{sI}}_{\mathbf{p},\mathbf{q}}$ & Vector pointing from source images to the receiver \\
        $\mathbf{R}^{\text{r}\to\text{sI}}_{\mathbf{p},\mathbf{q}}=-\mathbf{R}^{\text{sI}\to\text{r}}_{\mathbf{p},\mathbf{q}}$ & Vector pointing from receiver to source images \\
        $\mathbf{R}^{\text{s}\to\text{rI}}_{\tilde{\mathbf{p}},\tilde{\mathbf{q}}}$ & The reversed vector corresponding to the same reflection path of $\mathbf{R}^{\text{r}\to\text{sI}}_{\mathbf{p},\mathbf{q}}$ \\
		\bottomrule
	\end{tabular}
	\label{tab:tabparameters}
\end{table*}

The ISM was first applied to acoustics by Carslaw \cite{Carslaw1899} and later implemented using a digital computer by Allen and Berkley \cite{Allen1979} to model the impulse response between a point source and an omnidirectional receiver in a reverberant rectangular room. 

To derive the standard ISM, consider a rectangular room with length, width, and height given by $L_x, L_y$ and $L_z$, as well as an acoustic point source and an omidirectional receiver inside of the room at positions $\mathbf{x}_\text{s} = (x_\text{s},y_\text{s},z_\text{s})$ and $\mathbf{x}_\text{r} = (x_\text{r},y_\text{r},z_\text{r})$, given in Cartesian coordinates, respectively.
The RTF can be expressed as a weighted sum of free-space Green's functions attenuated by the walls, that is:
\begin{align}\label{eq:ism}
H_{\text{ISM}}(\mathbf{x}_\text{r} | \mathbf{x}_\text{s} , k) = \sum_{ \mathbf{p} \in \mathcal{P} } \sum_{ \mathbf{q} \in \mathcal{Q} } \beta(\mathbf{p},\mathbf{q}) G(\mathbf{x}_\text{r} | \mathbf{x}^{\text{sI}}_{\mathbf{p},\mathbf{q}} , k)~,
\end{align}
where $k$ is the wavenumber. The triples $\mathbf{p}$ and $\mathbf{q}$ determine the position of the images and the attenuation factor $\beta(\mathbf{p},\mathbf{q})$. The triple $\mathbf{p} = (p_x,p_y,p_z)$ is used to determine whether the image source is mirrored w.r.t. walls at $x=0$, $y=0$ or $z=0$, and its elements take values of $0$ or $1$, resulting in a set $\mathcal{P}$ of eight combinations. Each element of the other triple $\mathbf{q} = (q_x,q_y,q_z)$ takes a value between $[-N_m, N_m]$, and it includes higher order reflections, where $N_m$ is used to limit computational complexity and circular convolution errors~\cite{Jarrett2012b}, resulting in a set $\mathcal{Q}$ of $(2N_m+1)^3$ combinations. 
$\beta(\mathbf{p},\mathbf{q}) = \beta_{x_1}^{|q_x-p_x|} \beta_{x_2}^{|q_x|} \beta_{y_1}^{|q_y-p_y|} \beta_{y_2}^{|q_y|} \beta_{z_1}^{|q_z-p_z|} \beta_{z_2}^{|q_z|}$ and  $\beta_{x_1},\beta_{x_2},\beta_{y_1},\beta_{y_2},\beta_{z_1},\beta_{z_2}$ are the angle-independent reflection coefficients of the walls, and the subscripts $a_1,a_2$ ($a\in \{x,y,z\}$) of $\beta_{(\cdot)}$  correspond to the boundaries at $a=0$ and the boundaries at $a=L_a$, respectively. For an image source with position
\begin{equation}
    \mathbf{x}^{\text{sI}}_{\mathbf{p},\mathbf{q}} = 
			\begin{bmatrix}
				x_\text{s} - 2 p_x x_\text{s} +2 q_x L_x \\
				y_\text{s} - 2 p_y y_\text{s} +2 q_y L_y \\
				z_\text{s} - 2 p_z z_\text{s} +2 q_z L_z
			\end{bmatrix}~, \label{eq:sourceimage}
\end{equation}
the Green's function can be written as
\begin{equation}
    G(\mathbf{x}_\text{r} | \mathbf{x}^{\text{sI}}_{\mathbf{p},\mathbf{q}} , k) = \frac{e^{-i \, k || \mathbf{R}^{\text{sI}\to\text{r}}_{\mathbf{p},\mathbf{q}} ||}}{4\pi || \mathbf{R}^{\text{sI}\to\text{r}}_{\mathbf{p},\mathbf{q}}||}~,\label{eq:Greensfunction}
\end{equation}
where $i$ denotes the imaginary unit.
The vector pointing from every image of the source to the receiver is expressed as
\begin{equation}
    \mathbf{R}^{\text{sI}\to\text{r}}_{\mathbf{p},\mathbf{q}} = \mathbf{x}_\text{r}- \mathbf{x}^{\text{sI}}_{\mathbf{p},\mathbf{q}} = 
			\begin{bmatrix}
				x_\text{r} - x_\text{s} + 2p_x x_\text{s} - 2q_x L_x \\
				y_\text{r} - y_\text{s} + 2p_y y_\text{s} - 2q_y L_y \\
				z_\text{r} - z_\text{s} + 2p_z z_\text{s} - 2q_z L_z
			\end{bmatrix}~.\label{eq:sourceimagetoreceiver}
\end{equation}

\subsection{Extensions of the standard ISM}

Many extensions to the ISM have been proposed to facilitate more realistic modeling of room acoustics. In this section, we discuss two important extensions of the ISM  relevant to the present study, namely, using angle-dependent reflection coefficients and including source and receiver directivities.

\subsubsection{Angle-dependent reflection coefficients}

In Eq.~\eqref{eq:ism}, angle-independent reflection coefficients are used.
Aretz \textit{et al.} \cite{aretz2014application} have shown that including the angle of incidence when specifying the boundary conditions in the ISM can result in solutions that agree with FEM solutions. As such, angle-dependent reflection coefficients are used in this work.

For each reflection path represented via the indices $(\mathbf{p},\mathbf{q})$ in the ISM, three incident angles in Cartesian coordinates are needed, i.e., 
\begin{equation}
    \theta_{a}(\mathbf{p},\mathbf{q}) = \arccos \bigg( \frac{a_{\mathbf{R}^{\text{sI}\to\text{r}}_{\mathbf{p},\mathbf{q}}}}{\| \mathbf{R}^{\text{sI}\to\text{r}}_{\mathbf{p},\mathbf{q}} \|} \bigg)~, \label{eq:inc_angles}
\end{equation}
where $a\in \{x,y,z\}$, $a_{\mathbf{R}^{\text{sI}\to\text{r}}_{\mathbf{p},\mathbf{q}}}$ is the corresponding coordinate of the vector $\mathbf{R}^{\text{sI}\to\text{r}}_{\mathbf{p},\mathbf{q}}$ and $\theta_{a}(\mathbf{p},\mathbf{q})$ is the incident angle on the walls perpendicular to the $a-$axis. The incident angle $\theta_{a}(\mathbf{p},\mathbf{q})$ is the same for the reflections at the walls perpendicular to the same axis in a rectangular room \cite{aretz2014application}. For a uniform normalized impedance, $\zeta$, the angle-dependent reflection coefficients are given by~\cite{Kuttruff_2017}
\begin{align}
    \beta_{a_2} =&~\beta_{a_1} = \frac{\zeta \cos(\theta_a(\mathbf{p},\mathbf{q})) - 1}{\zeta \cos(\theta_a(\mathbf{p},\mathbf{q})) + 1}~. \label{eq:angdeprefcoef}
\end{align}
From this point onward, the attenuation term due to the reflections, $\beta(\mathbf{p},\mathbf{q})$, contains the angle-dependent reflection coefficients defined in Eq.~\eqref{eq:angdeprefcoef}. Note that, while the reflection properties of surfaces are generally frequency dependent, we use a frequency-independent impedance $\zeta$ for all walls in this work without loss of generality.

\subsubsection{Extension to the spherical harmonic domain}

In practice, the sound source and receiver in a room usually exhibit directional properties that are far more complex than those of a point source or an omnidirectional receiver. Various methods for incorporating source and receiver directivities into the ISM have been presented in the literature, e.g., \cite{Kompis1993, Habets2006, Wabnitz2010, Betlehem2012, Hafezi2015, Pyroomacoustics2018, Brinkmann2018, Gelderblom2021, Srivastava2022}. In this section, we present the Spherical Microphone array Impulse Response (SMIR) generator proposed by Jarrett \textit{et al.}~\cite{Jarrett2012a}.

The directivity of a source or receiver depends not only on the directional properties of the transducer itself but also on acoustic wave interaction with the physical object on which the transducer is mounted. The SMIR generator extends the ISM to the spherical harmonic domain, which facilitates the incorporation of the acoustic diffraction effect due to a rigid sphere and enables the simulation of RTFs for microphones mounted on such a sphere. When considering the receiver at $\mathbf{x}_{\text{r}}$ as the center of the spherical array with an omnidirectional microphone positioned at $\mathbf{y}^{({\text{r}})}$, the SMIR generator computes the RTF via
\begin{align}
H_{\text{SMIR}}(\mathbf{y}^{({\text{r}})} | \mathbf{x}_\text{s} , k) = \sum_{ \mathbf{p} \in \mathcal{P} } \sum_{ \mathbf{q} \in \mathcal{Q} } \beta(\mathbf{p},\mathbf{q})
	G_{\text{N}}(\mathbf{y}^{({\text{r}})} | \widetilde{\mathbf{R}}^{\text{r}\to\text{sI}}_{\mathbf{p},\mathbf{q}} , k)~, \label{eq:SMIR}
\end{align}
where $G_N(\cdot|\cdot)$ denotes the Green's function fulfilling Neumann boundary conditions on the sphere and $\widetilde{\mathbf{R}}^{\text{r}\to\text{sI}}_{\mathbf{p},\mathbf{q}}$ is the vector pointing from the center of the array to the source image in spherical coordinates. Detailed expressions of $H_{\text{SMIR}}(\mathbf{y}^{({\text{r}})} | \mathbf{x}_\text{s} , k)$ can be found in the literature \cite{Jarrett2012a}. It is worth noting that the Green's function given in Eq.~\eqref{eq:Greensfunction} and the Neumann Green's function in Eq.~\eqref{eq:SMIR} use different coordinate systems for the input arguments.

The acoustic diffraction on the sphere affects the directivity of the microphones on the sphere. Further directional properties of the source and receiver are usually implemented as a gain factor for each reflection path in the ISM. For example, Eq.~\eqref{eq:SMIR} is modified by considering the source directivity~\cite{Hafezi2015} yielding the expression: 
\begin{align}
H_{\text{SMIR+}}(\mathbf{y}^{({\text{r}})} | \mathbf{x}_\text{s} , k) =& \sum_{ \mathbf{p} \in \mathcal{P} } \sum_{ \mathbf{q} \in \mathcal{Q} } \beta(\mathbf{p},\mathbf{q}) g(\theta^{e,o}_{\mathbf{p},\mathbf{q}},\phi^{e,o}_{\mathbf{p},\mathbf{q}},k)\notag \\
&\times G_{\text{N}}(\mathbf{y}^{({\text{r}})} | \tilde{\mathbf{R}}^{\text{r}\to\text{sI}}_{\mathbf{p},\mathbf{q}} , k)~, \label{eq:SMIR+}
\end{align}
where $\theta^{e,o}_{\mathbf{p},\mathbf{q}},\phi^{e,o}_{\mathbf{p},\mathbf{q}}$ are the emission inclination and azimuth angles of the image source with indexes $(\mathbf{p},\mathbf{q})$ with respect to the orientation of the source directivity pattern. The angles $\theta^{e,o}_{\mathbf{p},\mathbf{q}},\phi^{e,o}_{\mathbf{p},\mathbf{q}}$ can be obtained by calculating the angles between the source orientation, e.g., the main direction of the directivity pattern at the cell with indexes $(\mathbf{p},\mathbf{q})$, and the vector from the source image with indexes $(\mathbf{p},\mathbf{q})$ to the receiver.
For first-order directivity patterns, $g(\cdot)$, can be expressed as~\cite{Hafezi2015}
\begin{equation}
    g(\theta^{e,o}_{\mathbf{p},\mathbf{q}},\phi^{e,o}_{\mathbf{p},\mathbf{q}},k) = \delta + (1 - \delta) \cos{\theta^{e,o}_{\mathbf{p},\mathbf{q}}} \sin{\phi^{e,o}_{\mathbf{p},\mathbf{q}}}~,
\end{equation}
where $ 0 \leq \delta \leq 1$. The relation between the emission and impinging angles, i.e., the mirrored emission angles after all reflections, depends on the number of reflections by walls perpendicular to different axes. A detailed formulation of this relation is provided by Hafezi \textit{et al.} \cite{Hafezi2015}. 
However, the phase information of the source directivity is missing in this approach. 

\subsection{Generalized image source method}
\label{sec:22}

\begin{figure}[t]
	\fig{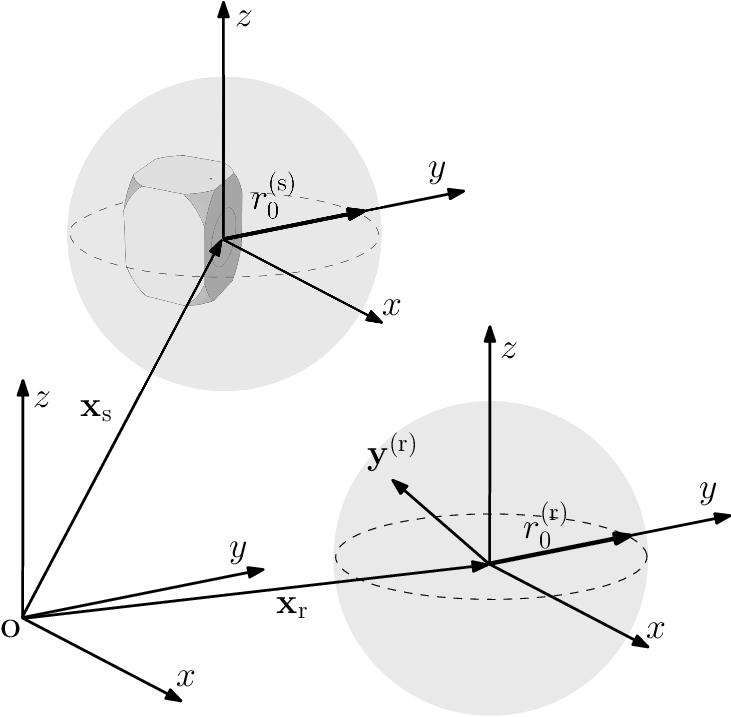}{0.5\textwidth}{}
	\caption{Illustration of the scenario modeled by Samarasinghe \emph{et al.}~\cite{samarasinghe2018spherical}. A directional source is positioned inside a rectangular room. The room transfer function from the source located at $\mathbf{x}_\text{s}$ to an observation point $\mathbf{y}^{(\text{r})}$ around the receiver located at $\mathbf{x}_\text{r}$ can be simulated.}
	\label{fig:GISM_formulation}
\end{figure}

The method introduced in Eq.~\eqref{eq:SMIR+} to model directional sources and receivers is restricted to directivity patterns with known analytical expressions. In practice, directivity patterns may exhibit complex shapes without analytical expressions. Samarasinghe \textit{et al.} \cite{samarasinghe2018spherical} further extended the ISM to facilitate the parameterization of the transfer function between a source with arbitrary directivity and a receiver region. The directional properties of the receiver region can be formed using beamforming techniques based on the received sound field. In this section, we describe the GISM, which is the foundation of our proposed method. 

The GISM decomposes the transfer function into three parts: i)~an outgoing sound field from the directional source, ii)~an incident sound field captured by the receiver that can be directional, and iii)~the mode coupling coefficients that couple the source and receiver directivities in the spherical harmonic domain. 

Consider an outgoing sound field from a directional source, for example a loudspeaker, located at $\mathbf{x}_{\text{s}}$ as shown in Fig.~\ref{fig:GISM_formulation}. The sound field observed at an arbitrary point $\mathbf{r}$ outside of the sphere enclosing the entire speaker can be computed as~\cite{Williams1999}: 
\begin{equation}
	P (\mathbf{r},k) = \sum_{n=0}^\infty \sum_{m=-n}^{n} C^{(\text{s})}_{n,m}(k) h_n (k d) Y_{n,m} (\theta,\phi)~, \label{eq:GISMoutgoing}
\end{equation} 
where $(d,\theta,\phi)$ denote the spherical coordinates associated with $\mathbf{r}$, viz. the distance $d$, the inclination angle $\theta$ and the azimuth angle $\phi$. Moreover, $h_n(\cdot)$ denotes the spherical Hankel function of order $n$, $Y_{n,m}$ denotes the spherical harmonic function of order $n$ and mode $m$, and $C^{(\text{s})}_{n,m}(k)$ represent the spherical harmonic source directivity coefficients. In practice, the infinite summation in Eq.\eqref{eq:GISMoutgoing} is truncated to a maximum order $N = \left \lceil k r_0^{(\text{s})} \right \rceil$ where $r_0^{(\text{s})}$ denotes the radius of the sphere that is large enough to surround the loudspeaker \cite{Zhang2010, Samarasinghe2016}.

The room transfer function from the source $\mathbf{x}_{\text{s}}$ to any point $\mathbf{y}^{(\text{r})}$ with spherical coordinates $(d_y^{({\text{r}})},\theta_y^{({\text{r}})},\phi_y^{({\text{r}})})$ relative to $\mathbf{x}_{\text{r}}$ can be expressed using the GISM as~\cite{samarasinghe2018spherical}: 
\begin{align}
	H_{\text{GISM}}(\mathbf{y}^{({\text{r}})} | \mathbf{x}_\text{s} , k) =& \sum_{v=0}^V \sum_{u=-v}^{v} \sum_{n=0}^N \sum_{m=-n}^{n} C^{(\text{s})}_{n,m}(k) \gamma_{v,u}^{n,m}(k)\notag \\
	&\times j_v (k d_y^{(\text{r})}) Y_{v,u} (\theta_y^{(\text{r})},\phi_y^{(\text{r})})~, \label{eq:GISMRTF}
\end{align}
where $V = \left \lceil k \, d_y^{(r)} \right \rceil$ denotes the truncation limit that produces a small error \cite{Ward2001} and $j_v(\cdot)$ denotes the $n$th-order spherical Bessel function of the first kind. The reverberant mode coupling coefficients $\gamma_{v,u}^{n,m}(k)$ are given by
\begin{align}
    \gamma_{v,u}^{n,m}(k) =& \sum_{ \mathbf{p} \in \mathcal{P} } \sum_{ \mathbf{q} \in \mathcal{Q} } \beta(\mathbf{p},\mathbf{q}) (-1)^{(p_y + p_z)m + p_z n} \notag \\ 
    &\times \alpha_{v,u}^{n,(-1)^{p_x +p_y} m}(\mathbf{R}^{\text{sI}\to\text{r}}_{\mathbf{p},\mathbf{q}} )~,\label{eq:GISMreverbmodecoupling}
\end{align}  
where $\mathbf{R}^{\text{sI}\to\text{r}}_{\mathbf{p},\mathbf{q}}$ has spherical coordinates $(||\mathbf{R}^{\text{sI}\to\text{r}}_{\mathbf{p},\mathbf{q}}||,\Omega^{\text{sI}\to\text{r}}_{\mathbf{p},\mathbf{q}})$, and $\Omega^{\text{sI}\to\text{r}}_{\mathbf{p},\mathbf{q}} = (\theta^{\text{sI}\to\text{r}}_{\mathbf{p},\mathbf{q}},\phi^{\text{sI}\to\text{r}}_{\mathbf{p},\mathbf{q}})$. 
The additional factors $(-1)^{(p_y + p_z)m + p_z n}$ and $(-1)^{p_x +p_y}$ result from the mirror effect of the image sources~\cite{rafaely2015fundamentals} on the spherical harmonics, which is a more general description compared to the use of emission and impinging angles as applied in the SMIR+ method~\cite{Hafezi2015}.
Using $\mathbf{x}_\mathbf{0}$ with spherical coordinates $(d_{x_0},\theta_{x_0},\phi_{x_0})$ instead of $\mathbf{R}^{\text{sI}\to\text{r}}_{\mathbf{p},\mathbf{q}}$ for brevity, the single-path mode-coupling coefficients $\alpha_{v,u}^{n,m}(\cdot)$ in Eq.\eqref{eq:GISMreverbmodecoupling} can be derived using the translation of fields \cite{rafaely2015fundamentals},
\begin{align}
	\alpha_{v,u}^{n,m}(\mathbf{x}_\mathbf{o})=&
	4 \pi i^{v-n} (-1)^{m} \displaystyle \sum_{l=|n-v|}^{n+v} i^l  h^{(2)}_l(kd_{x_0}) \notag \\
	&\times Y_{l,m-u}(\theta_{x_0},\phi_{x_0})W_1 W_2 \xi	~, \label{eq:GISMmodecoupling}
\end{align}
with $W_1$ and $W_2$ denoting Wigner 3 - j symbols,
\begin{align}
	W_1 = \begin{pmatrix}
		n & v & l \\
		0 & 0 & 0 
	\end{pmatrix} 
	~ , ~
	W_2 = \begin{pmatrix}
		n & v & l \\
		-m & u & m-u 
	\end{pmatrix}~, \label{eq:Wigners}
\end{align}
and $\xi = \sqrt{(2n +1)(2v+1)(2l+1)/4\pi}$. 

\subsection{Fast source room receiver method}
\label{sec:24}

The Fast Source Room Receiver (FSRR) method, proposed by Luo and Kim~\cite{luo2021fast}, also simulates the RTF in the spherical harmonic domain. The RTF parameterization includes an approximation of the spherical harmonic description of the source and receiver directivities and utilizes a simplification of the cross-modal coupling, i.e., the mode coupling coefficients in Eq.~\eqref{eq:GISMreverbmodecoupling}. The method also uses a randomized sign parameter. Notably, the FSRR method uses reversed reflection paths - a technique explored in this work for deriving a simplified version of the proposed method, which is presented in Sec.~\ref{sec:34}.

 Using our notation, the FSRR method RTF is given by
\begin{align}
    H_{\text{FSRR}} & (\mathbf{x}_\text{r} | \mathbf{x}_\text{s}, k) = \sum_{ \mathbf{p} \in \mathcal{P} } \sum_{ \mathbf{q} \in \mathcal{Q} } B_{\mathbf{p},\mathbf{q}} M_{\mathbf{p},\mathbf{q}} \frac{e^{-i k ||\mathbf{R}^{\text{r}\to\text{sI}}_{\mathbf{p},\mathbf{q}}|| }}{||\mathbf{R}^{\text{r}\to\text{sI}}_{\mathbf{p},\mathbf{q}}||}  \notag \\
    &\times \sum_{n=0}^N \sum_{m=-n}^{n} \tilde{C}^{(\text{s})}_{n,m}(k) Y_{n,m}(\Omega^{\text{s}\to\text{rI}}_{\tilde{\mathbf{p}},\tilde{\mathbf{q}}}) \notag \\
    &\times \sum_{v=0}^V \sum_{u=-v}^{v} \tilde{C}^{(\text{r})}_{v,u}(k) Y_{v,u}(\Omega^{\text{r}\to\text{sI}}_{\mathbf{p},\mathbf{q}})~, \label{eq:FSRRRTF}
\end{align}
where $B_{\mathbf{p},\mathbf{q}}$ randomly takes value between $1$ and $-1$, $M_{\mathbf{p},\mathbf{q}}$ is a function fitted to frequency-dependent absorption coefficients, $\tilde{C}$ is a directivity coefficient obtained at 1~m from either the source or the receiver. The modified triples $\tilde{\mathbf{p}},\tilde{\mathbf {q}}$ are defined in Sec.~\ref{sec:34} (Eq.~\eqref{eq:modifiedtriples}), $\Omega^{\text{sI} \to \text{r}}_{ \tilde{ \mathbf{p}},\tilde{\mathbf{q}}}$ is the direction of a vector pointing from a source image to the receiver, and $\Omega^{\text{r}\to\text{sI}}_{\tilde{\mathbf{p}},\tilde{\mathbf{q}}}$ is the direction of the vector corresponding to the reversed path of the same reflection. Definitions of the two vectors are given in Sec.~\ref{sec:34}. 

\section{Proposed method}
\label{sec:3}

We have seen that the SMIR generator relies on an analytical descriptions of directivity, the GISM simulates either a directive source or a directive receiver, and the FSRR method approximates the RTF between directional sources and receivers. In this section, we propose a method for simulating RTFs that can include near-field diffraction effects of both directional sources and directional receivers.

\subsection{Problem formulation}
\label{sec:31}

\begin{figure}[t]
	\centering
	\fig{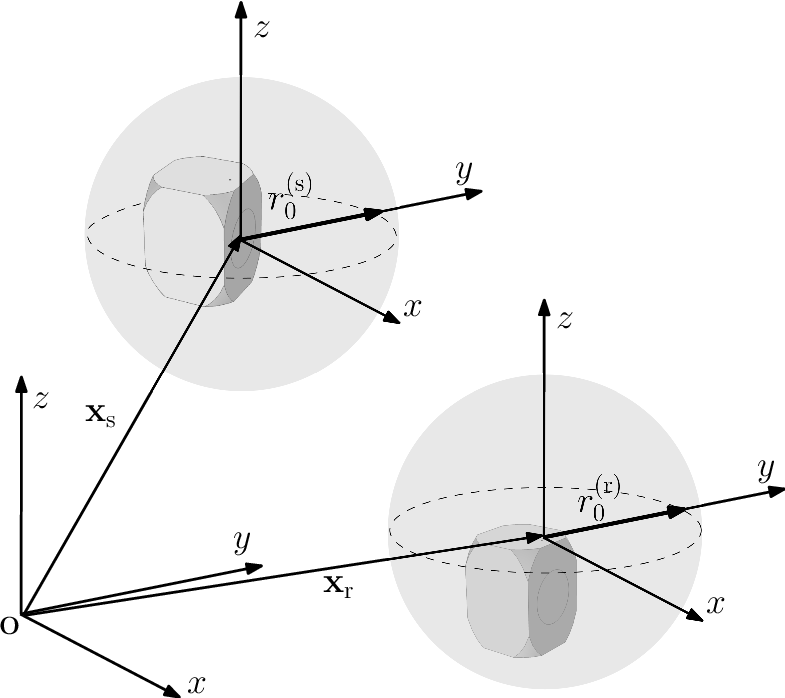}{0.5\textwidth}{}
	\caption{Illustration of the scenario considered for the proposed method. Two speakers are positioned inside a rectangular room. The source speaker has a vibrating piston on one of its surfaces located at $\mathbf{x}_\text{s}$. The receiver speaker has a microphone on the top surface located at $\mathbf{x}_\text{r}$. The room transfer function from $\mathbf{x}_\text{s}$ to $\mathbf{x}_\text{r}$ is of interest. Not shown here is the room in which the speakers are placed. The two spherical regions marked in light gray are nonoverlapping.}
	\label{fig:DEISM_formulation}
\end{figure}

We consider the following scenario: one source device and one receiver device are positioned in a rectangular room, as depicted in Fig.~\ref{fig:DEISM_formulation}. The source contains a sound-radiating transducer located at $\mathbf{x}_{\text{s}}$, e.g., a vibrating piston on one surface, and a microphone located at $\mathbf{x}_{\text{r}}$ is mounted on the receiver. The two devices can have arbitrary shapes and materials, and the transducers can exhibit arbitrary directivity patterns. The acoustic field generated by the source transducer interacts with both devices and the walls and is captured by the microphone. We aim to incorporate the local diffraction effects around the devices into the RTFs based on the ISM. The existing methods are unable to resolve the problem under consideration due to the following reasons:
\begin{itemize}
    \item As shown by Xu \textit{et al.}\cite{Xu_et_al_2022}, the terms $j_v (k d_y^{(\text{r})}) Y_{v,u} (\theta_y^{(\text{r})},\phi_y^{(\text{r})})$ corresponding to the incident sound field in Eq.~\eqref{eq:GISMRTF} cannot model the diffraction around the receiver. Therefore, even utilizing the acoustic reciprocity principle, the GISM can only model the RTFs when either the source or the receiver does not include local diffraction effects. 
    \item The FSRR method~\cite{luo2021fast} extends the ISM to include the directivity of both the source and receiver using a far-field approximation and, therefore, cannot model near-field directivities.
\end{itemize}

\subsection{Source and receiver directivities}
\label{sec:32}

The directivity of a transducer mounted on a device can either be numerically simulated or practically measured at discrete points on a sphere around a local origin. When sound waves are generated or captured by the mounted transducers, e.g., a circular piston at $\mathbf{x}_\text{s}$ or a microphone at $\mathbf{x}_\text{r}$ as shown in Fig.~\ref{fig:DEISM_formulation}, acoustic diffraction occurs due to the wave interaction with the speakers. As shown in previous works~\cite{Williams1999,Ahrens2021b,Xu_et_al_2022}, the total sound radiation of a source device can be quantified using directivity coefficients defined on a transparent sphere with radius $r_0^{(\text{s})}$. Using reciprocity, the receiver directivity can be obtained analogously to the source directivity~\cite{Xu_et_al_2022}. As an example, consider the source device. If the sound field is sampled at $J$ discrete points on the transparent sphere, one can reformulate Eq.~\eqref{eq:GISMoutgoing} with truncated order $N$ as 
\begin{equation}
	P (\mathbf{r}_j,k) = \sum_{n=0}^N \sum_{m=-n}^{n} P_{n,m}(k,r_0^{(\text{s})}) Y_{n,m} (\theta_j,\phi_j)~, \label{eq:sampledJpoints}
\end{equation} 
where $\mathbf{r}_j$ denotes the the $j$th position on the transparent sphere, $(\theta_j,\phi_j)$ the corresponding direction, $P_{n,m}(r_0^{(\text{s})},k) = C^{(\text{s})}_{n,m}(k) h_n (k r_0^{(\text{s})})$ is the spherical wave spectrum \cite{Williams1999} and $j \in \{1,2,\ldots,J\}$. Different choices of the distribution of $(\theta_j,\phi_j)$ can be found in the literature \cite{rafaely2015fundamentals}. If the number of samples is larger than $(N+1)^2$, the spherical wave spectrum can be calculated from the sampled sound field by solving an over-determined linear system using a least-squares approach. The directivity coefficients of the source can then be determined by $C^{(\text{s})}_{n,m}(k) = P_{n,m}(r_0^{(\text{s})},k) / h_n (k \, r_0^{(\text{s})})$. The receiver directivity coefficients $C^{(\text{r})}_{n,m}(k)$ can be obtained following the same procedure by making use of reciprocity~\cite{Xu_et_al_2022}.

\subsection{Room transfer function}
\label{sec:33}

The following parameterization of the transfer function between two transducers with arbitrary directivities in the free field, $H^{(\text{F})}$, was proposed by Xu \emph{et al.}~\cite{Xu_et_al_2022}:
\begin{align}
	H^{(\text{F})} (\mathbf{x}_\text{r} | \mathbf{x}_\text{s}, k) =& \sum_{v=0}^V \sum_{u=-v}^{v} \sum_{n=0}^N \sum_{m=-n}^{n} C^{(\text{s})}_{n,m}(k) \notag \\ &\times \alpha_{v,u}^{n,m}(k) D^{(\text{r})}_{v,u}(k)~, \label{eq:DEISMRTFfreefield}
\end{align}
where $\alpha_{v,u}^{n,m}(k)$ represent the free-field mode coupling coefficients. Note that the two transparent spheres shown in Fig.~\ref{fig:DEISM_formulation} are nonoverlapping in this work. However, when both the source and receiver are on the same device, one can measure or simulate the direct path to avoid the overlapping of the two transparent spheres. The following relation between $D^{(\text{r})}_{v,u}(k)$ and the directivity coefficients $C^{(\text{r})}_{v,u}(k)$ obtained via reciprocity for the receiver was derived in~\cite{Xu_et_al_2022}
\begin{equation}
    D_{v,u}^{(\text{r})}(k) = \frac{i (-1)^u}{k} C_{v,-u}^{(\text{r})}(k)~.\label{eq:relation_D_C}
\end{equation}
Since the summation over source images is a linear operation and the mirroring effects are already included in $\gamma_{v,u}^{n,m}$ in Eq.~\eqref{eq:GISMreverbmodecoupling}, the total RTF can be written by substituting $\gamma_{v,u}^{n,m}(k)$ for $\alpha_{v,u}^{n,m}(k)$ in Eq.~\eqref{eq:DEISMRTFfreefield}, yielding:
\begin{align}
	H_{\text{DEISM}} (\mathbf{x}_\text{r} | \mathbf{x}_\text{s}, k) =& \sum_{v=0}^V \sum_{u=-v}^{v} \sum_{n=0}^N \sum_{m=-n}^{n} C^{(\text{s})}_{n,m}(k) \notag \\ &\times \gamma_{v,u}^{n,m}(k) D^{(\text{r})}_{v,u}(k)~.\label{eq:DEISMRTFreverb1}
\end{align}
The modified reverberant mode coupling coefficients can be defined, based on Eq.~\eqref{eq:GISMreverbmodecoupling} and the reciprocal relation in Eq.~\eqref{eq:relation_D_C}, as follows:
\begin{equation}
    \tilde{\alpha}_{v,u}^{n,m}(k) = \frac{i (-1)^u}{k} \gamma_{v,u}^{n,m}(k) ~. \label{eq:DEISMmodecoupling}
\end{equation}  
The final proposed RTF is therefore expressed as 
\begin{align}
	H_{\text{DEISM}} (\mathbf{x}_\text{r} | \mathbf{x}_\text{s}, k) =& \sum_{v=0}^V \sum_{u=-v}^{v} \sum_{n=0}^N \sum_{m=-n}^{n} C^{(\text{s})}_{n,m}(k) \notag \\ &\times \tilde{\alpha}_{v,u}^{n,m}(k) C^{(\text{r})}_{v,-u}(k)~, \label{eq:DEISMRTFreverbfinal}
\end{align}
where, for clarity, the proposed method is identified using the acronym Diffraction Enhanced Image Source Method (DEISM).

The GISM RTF, i.e., the simulation of a directional source to a local receiving region, can be recovered from the proposed RTF as follows.
Similarly to Xu \emph{et al.}~\cite{Xu_et_al_2022}, for a receiver region without any physical objects and an observation point $\mathbf{y}^{(\text{r})}$ with spherical coordinates $(d_y^{(\text{r})},\theta_y^{(\text{r})},\phi_y^{(\text{r})})$ relative to $\mathbf{x}_\text{r}$, as shown in Fig.~\ref{fig:GISM_formulation}, the receiver directivity coefficients can be written as
\begin{equation}
    C_{v,u}^{(\text{r})} = - i k j_v(k d_y^{(\text{r})}) Y_{v,u}^*(\theta_y^{(\text{r})},\phi_y^{(\text{r})})~. \label{eq:monopolereceiverC}
\end{equation}
By substituting Eq.~\eqref{eq:monopolereceiverC} into Eq.~\eqref{eq:DEISMRTFreverbfinal}, and utilizing the property of spherical harmonics $Y_{v,u}^*(\theta_y^{(\text{r})},\phi_y^{(\text{r})}) = (-1)^u Y_{v,-u}(\theta_y^{(\text{r})},\phi_y^{(\text{r})})$, one can obtain the same expression of the RTFs as the GISM shown in Eq.~\eqref{eq:GISMRTF}.

\subsection{Far-field approximation}
\label{sec:34}

\begin{figure*}[t]
    \fig{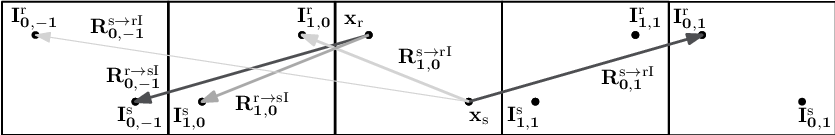}{\textwidth}{}
    \caption{An example of reversed reflection paths along the $x-$direction used in the ISM. For an odd number of reflections, $\mathbf{R}^{\text{r}\to\text{sI}}_{\mathbf{1},\mathbf{0}}$ and $\mathbf{R}^{\text{s}\to\text{rI}}_{\mathbf{1},\mathbf{0}}$ share the same-length, but reversed reflection path. However, for an even number of reflections, e.g., $\mathbf{R}^{\text{r}\to\text{sI}}_{\mathbf{0},\mathbf{-1}}$, the reversed path $\mathbf{R}^{\text{s}\to\text{rI}}_{\mathbf{0},\mathbf{1}}$ is symmetric w.r.t the original room cell instead of $\mathbf{R}^{\text{s}\to\text{rI}}_{\mathbf{0},\mathbf{-1}}$.} 
    \label{fig:reciprocalimages}
\end{figure*}

The computational cost of the DEISM RTF is expected to increase with frequency since the directivity patterns tend to be more complicated at higher frequencies \cite{Tylka2015a}, resulting in higher orders of the directivities, i.e., larger $N$ and $V$, required in the DEISM. As either $N$ or $V$ increases, the number of summations over $l$ in the mode coupling coefficients in Eq.~\eqref{eq:GISMreverbmodecoupling} increases, thus consuming more computational resources. Therefore, it is necessary to investigate the possibility of simplifying the DEISM in Eq.~\eqref{eq:DEISMRTFreverbfinal} while maintaining sufficient accuracy. To this end, a far-field approximation is applied to each reflection path of the RTF in Eq.~\eqref{eq:DEISMRTFreverbfinal} in this section.

Without loss of generality, we consider a reflection path from the source image to the receiver, denoted by the vector  $\mathbf{R}^{\text{sI}\to\text{r}}_{\mathbf{p},\mathbf{q}}$, for the following analysis. The mode coupling coefficients from Eq.~\eqref{eq:DEISMmodecoupling} for a single reflection path can be expressed as follows, 
\begin{align}
    \tilde{\alpha}_{v,u}^{n,m}(k) =& \beta(\mathbf{p},\mathbf{q}) \varLambda(\mathbf{p},m,n) \frac{4 \pi i}{k}  i^{v-n} \\
    &\times  (-1)^{m^\prime+u} \displaystyle \sum_{l=|n-v|}^{n+v} i^l  h^{(2)}_l(k||\mathbf{R}^{\text{sI}\to\text{r}}_{\mathbf{p},\mathbf{q}}||) \notag \\
	&\times Y_{l,m^\prime-u}(\Omega^{\text{sI}\to\text{r}}_{\mathbf{p},\mathbf{q}})W_1 W_2 \xi	~,
\end{align}
where $m^\prime = (-1)^{p_x + p_y}m$ and $\varLambda(\mathbf{p},m,n) = (-1)^{(p_y + p_z)m + p_z n}$ are the additional effect of the mirroring on the spherical harmonics \cite{samarasinghe2018spherical}. 
Assuming $||\mathbf{R}^{\text{sI}\to\text{r}}_{\mathbf{p},\mathbf{q}}||$ is large, we employ the large argument of the spherical Hankel function, i.e., $h^{(2)}_l(k||\mathbf{R}^{\text{sI}\to\text{r}}_{\mathbf{p},\mathbf{q}}||) \approx i^{l+1}e^{-ik||\mathbf{R}^{\text{sI}\to\text{r}}_{\mathbf{p},\mathbf{q}}||}/k||\mathbf{R}^{\text{sI}\to\text{r}}_{\mathbf{p},\mathbf{q}}||$.
The modified mode coupling coefficients in far field can be expressed as
\begin{align}
    \tilde{\alpha}_{v,u}^{n,m}(k) \approx &  \beta(\mathbf{p},\mathbf{q}) \varLambda(\mathbf{p},m,n) \frac{4 \pi}{k}  i^{v-n} \notag\\
    &\times (-1)^{m^\prime+u+1} \frac{e^{-ik||\mathbf{R}^{\text{sI}\to\text{r}}_{\mathbf{p},\mathbf{q}}||}}{k||\mathbf{R}^{\text{sI}\to\text{r}}_{\mathbf{p},\mathbf{q}}||} \displaystyle \sum_{l=|n-v|}^{n+v} (-1)^l  \notag \\
	&\times Y_{l,m^\prime-u}(\Omega_{\mathbf{p},\mathbf{q}})W_1 W_2 \xi~.\label{eq:DEISMmodecouplingfarfield}
\end{align}
Note that the product of two spherical harmonics can be expressed using the formula~\cite{Edmonds2016}, 
\begin{align}
    & Y_{n,m}(\theta,\phi) Y_{v,-u}(\theta,\phi) = (-1)^{n+m+v-u}   \notag\\
    &\times \displaystyle \sum_{l=|n-v|}^{n+v} (-1)^l Y_{l,m-u}(\theta,\phi)W_1 W_2 \xi ~.\label{eq:producttwoSHs}
\end{align}
By applying Eq.~\eqref{eq:producttwoSHs} to Eq.~\eqref{eq:DEISMmodecouplingfarfield}, we can express the mode coupling coefficients as follows,
\begin{align}
    \tilde{\alpha}_{v,u}^{n,m}(k) \approx & \beta(\mathbf{p},\mathbf{q}) \varLambda(\mathbf{p},m,n) \frac{4 \pi}{k}  i^{v-n}  \notag \\
    & \times (-1)^{n+v-1} \frac{e^{-ik||\mathbf{R}^{\text{sI}\to\text{r}}_{\mathbf{p},\mathbf{q}}||}}{k||\mathbf{R}^{\text{sI}\to\text{r}}_{\mathbf{p},\mathbf{q}}||} \notag\\ 
    & \times Y_{n,m}(\Omega^{\text{sI}\to\text{r}}_{\mathbf{p},\mathbf{q}}) Y_{v,-u}(\Omega^{\text{sI}\to\text{r}}_{\mathbf{p},\mathbf{q}})~.
\end{align}
The transfer function of one reflection path using the far-field approximation model can thus be written as,
\begin{align}
	& H^{\mathbf{p},\mathbf{q}}_{\text{DEISM-LC}} (\mathbf{x}_\text{r} | \mathbf{x}_\text{s}, k) \approx -\beta(\mathbf{p},\mathbf{q})\frac{4 \pi}{k} \frac{e^{-ik||\mathbf{R}^{\text{sI}\to\text{r}}_{\mathbf{p},\mathbf{q}}||}}{k||\mathbf{R}^{\text{sI}\to\text{r}}_{\mathbf{p},\mathbf{q}}||}   \notag \\ 
	&\times \sum_{n=0}^N \sum_{m=-n}^{n}\varLambda(\mathbf{p},m,n) i^{-n} (-1)^{n} C^{(\text{s})}_{n,m}(k) Y_{n,m^\prime}(\Omega^{\text{sI}\to\text{r}}_{\mathbf{p},\mathbf{q}}) \notag \\
	&\times \sum_{v=0}^V \sum_{u=-v}^{v} i^{v} (-1)^{v} C^{(\text{r})}_{v,-u}(k) Y_{v,-u}(\Omega^{\text{sI}\to\text{r}}_{\mathbf{p},\mathbf{q}}) ~, \label{eq:DEISMRTFpathpq}
\end{align}
where DEISM-LC is used to refer to the Low-Complexity (LC) solution.

The terms $\varLambda(\mathbf{p},m,n)Y_{n,m^\prime}(\Omega^{\text{sI}\to\text{r}}_{\mathbf{p},\mathbf{q}})$ are a result of mirroring of spherical harmonics. 
Since the source images are used in Eq.~\eqref{eq:DEISMRTFpathpq}, the additional factors $\varLambda(\mathbf{p},m,n)$ and $m^\prime$ need to be associated with the source spherical harmonics. To further simplify the expressions in Eq.~\eqref{eq:DEISMRTFpathpq}, we propose to replace $\varLambda(\mathbf{p},m,n)Y_{n,m^\prime}(\Omega^{\text{sI}\to\text{r}}_{\mathbf{p},\mathbf{q}})$ by a simplified expression, which can be achieved as shown in the following by using the direction that corresponds to the reversed traveling path of the vector $\mathbf{R}^{\text{sI}\to\text{r}}_{\mathbf{p},\mathbf{q}}$. 

First, we note that the vector pointing from the receiver to the source image $\mathbf{R}^{\text{r}\to\text{sI}}_{\mathbf{p},\mathbf{q}}$ is the inverse of the vector $\mathbf{R}^{\text{sI}\to\text{r}}_{\mathbf{p},\mathbf{q}}$ given in Eq.~\eqref{eq:sourceimagetoreceiver}. In addition, we define the vector pointing from the source to the receiver images $\mathbf{R}^{\text{s}\to\text{rI}}_{\mathbf{p},\mathbf{q}}$, which can be expressed similarly to Eq.~\eqref{eq:sourceimagetoreceiver} as
\begin{align}
   \mathbf{R}^{\text{s}\to\text{rI}}_{\mathbf{p},\mathbf{q}} =& 
			\begin{bmatrix}
				x_\text{r} - x_\text{s} - 2p_x x_\text{r} + 2q_x L_x \\
				y_\text{r} - y_\text{s} - 2p_y y_\text{r} + 2q_y L_y \\
				z_\text{r} - z_\text{s} - 2p_z z_\text{r} + 2q_z L_z
			\end{bmatrix}~. \label{eq:sourcetoreceiverimage}
\end{align}
In spherical coordinates, the vectors $\mathbf{R}^{\text{r}\to\text{sI}}_{\mathbf{p},\mathbf{q}}$ and $\mathbf{R}^{\text{s}\to\text{rI}}_{\mathbf{p},\mathbf{q}}$ have directions $\Omega^{\text{r}\to\text{sI}}_{\mathbf{p},\mathbf{q}}$ and $\Omega^{\text{s}\to\text{rI}}_{\mathbf{p},\mathbf{q}}$, respectively. These two vectors are reversed reflection paths only when there are an odd number of reflections between parallel walls, as shown by the gray arrows in Fig.~\ref{fig:reciprocalimages}. To take the even number of reflections into account, as shown by the black arrows in Fig.~\ref{fig:reciprocalimages}, we use the modified vector $\mathbf{R}^{\text{s}\to\text{rI}}_{\tilde{\mathbf{p}},\tilde{\mathbf{q}}}$ (with directions $\Omega^{\text{s}\to\text{rI}}_{\tilde{\mathbf{p}},\tilde{\mathbf{q}}}$) that represents the reversed path of $\mathbf{R}^{\text{r}\to\text{sI}}_{\mathbf{p},\mathbf{q}}$ with its subscript indices $\tilde{\mathbf{p}},\tilde{\mathbf{q}} = (\tilde{p}_x,\tilde{p}_y,\tilde{p}_z),(\tilde{q}_x,\tilde{q}_y,\tilde{q}_z) $ taking the following expressions
\begin{align}
   \tilde{p}_a, \tilde{q}_a = 
   \begin{cases}
   p_a,q_a &\text{if } |2q_a-p_a|\mod 2 = 1 \\
   p_a^\prime,q_a^\prime &\text{if } |2q_a-p_a|\mod 2 = 0
   \end{cases}~,\label{eq:modifiedtriples}
\end{align}
where $a\in \{ x,y,z \}$, and $p_a^\prime,q_a^\prime$ represent the image indices that are symmetric to $p_a,q_a$ w.r.t the image indices $p_a,q_a=0,0$. The relation between $p_a^\prime,q_a^\prime$ and $p_a,q_a$ can be summarized as 
\begin{align}
    2q_a - p_a &= p_a^\prime - 2 q_a^\prime \\
    q_a^\prime &= \Big\lfloor \frac{p_a - 2q_a}{2} \Big\rfloor~,
\end{align}
where $\lfloor \cdot \rfloor$ denotes the floored division. 

Secondly, due to the properties of the mirroring effect on the spherical harmonics, one can write the following relation between the spherical harmonics associated with the spherical direction $\Omega^{\text{sI}\to\text{r}}_{\mathbf{p},\mathbf{q}}$ in Eq.~\eqref{eq:DEISMRTFpathpq} and $\Omega^{\text{s}\to\text{rI}}_{\tilde{\mathbf{p}},\tilde{\mathbf{q}}}$ introduced above
\begin{equation}
    \varLambda(\mathbf{p},m,n) Y_{n,m^\prime}(\Omega^{\text{sI}\to\text{r}}_{\mathbf{p},\mathbf{q}}) = Y_{n,m}(\Omega^{\text{s}\to\text{rI}}_{\tilde{\mathbf{p}},\tilde{\mathbf{q}}})~.\label{eq:mirrorrelation}
\end{equation}
Since $\mathbf{R}^{\text{r}\to\text{sI}}_{\mathbf{p},\mathbf{q}} = - \mathbf{R}^{\text{sI}\to\text{r}}_{\mathbf{p},\mathbf{q}}$, i.e., $\Omega^{\text{sI}\to\text{r}}_{\mathbf{p},\mathbf{q}}$ and $\Omega^{\text{r}\to\text{sI}}_{\mathbf{p},\mathbf{q}}$ have opposite directions, the following relation is obtained
\begin{equation}
    Y_{v,-u}(\Omega^{\text{sI}\to\text{r}}_{\mathbf{p},\mathbf{q}}) = (-1)^v Y_{v,-u}(\Omega^{\text{r}\to\text{sI}}_{\mathbf{p},\mathbf{q}})~.\label{eq:oppositerelation}
\end{equation}
Finally, by substituting Eqs.~\eqref{eq:mirrorrelation} and \eqref{eq:oppositerelation} in the approximated solution in Eq.~\eqref{eq:DEISMRTFpathpq}, we obtain the simplified expression
\begin{align}
	& H^{\mathbf{p},\mathbf{q}}_{\text{DEISM-LC}} (\mathbf{x}_\text{r} | \mathbf{x}_\text{s}, k) \approx -\beta(\mathbf{p},\mathbf{q})\frac{4 \pi}{k} \frac{e^{-ik||\mathbf{R}^{\text{r}\to\text{sI}}_{\mathbf{p},\mathbf{q}}||}}{k||\mathbf{R}^{\text{r}\to\text{sI}}_{\mathbf{p},\mathbf{q}}||}   \notag \\ 
	&\times \sum_{n=0}^N \sum_{m=-n}^{n} i^n C^{(\text{s})}_{n,m}(k) Y_{n,m}(\Omega^{\text{s}\to\text{rI}}_{\tilde{\mathbf{p}},\tilde{\mathbf{q}}}) \notag \\
	&\times \sum_{v=0}^V \sum_{u=-v}^{v} i^{v} C^{(\text{r})}_{v,u}(k) Y_{v,u}(\Omega^{\text{r}\to\text{sI}}_{\mathbf{p},\mathbf{q}}) ~. \label{eq:DEISMRTFpathpq2}
\end{align}
The RTF of the DEISM-LC is then given by
\begin{align}
	& H_{\text{DEISM-LC}} (\mathbf{x}_\text{r} | \mathbf{x}_\text{s}, k) = \sum_{ \mathbf{p} \in \mathcal{P} } \sum_{ \mathbf{q} \in \mathcal{Q} } -\beta(\mathbf{p},\mathbf{q})\frac{4 \pi}{k} \frac{e^{-ik||\mathbf{R}^{\text{r}\to\text{sI}}_{\mathbf{p},\mathbf{q}}||}}{k||\mathbf{R}^{\text{r}\to\text{sI}}_{\mathbf{p},\mathbf{q}}||}   \notag \\ 
	&\times \sum_{n=0}^N \sum_{m=-n}^{n} i^n C^{(\text{s})}_{n,m}(k) Y_{n,m}(\Omega^{\text{s}\to\text{rI}}_{\tilde{\mathbf{p}},\tilde{\mathbf{q}}}) \notag \\
	&\times \sum_{v=0}^V \sum_{u=-v}^{v} i^{v} C^{(\text{r})}_{v,u}(k) Y_{v,u}(\Omega^{\text{r}\to\text{sI}}_{\mathbf{p},\mathbf{q}}) ~. \label{eq:DEISMLCRTF}
\end{align}
The DEISM-LC is more computationally efficient since it does not require the summation over $l$ in the mode coupling coefficient in Eq.~\eqref{eq:GISMreverbmodecoupling}. The DEISM in Eq.~\eqref{eq:DEISMRTFreverbfinal} has a time complexity of $\mathcal{O}(IN^2V^2(N+V))$ per wavenumber $k$, where $I$ is the number of images, $N$ and $V$ are the maximum SH order used to describe directivities of the source and receiver. However, the time complexity of DEISM-LC is reduced to $\mathcal{O}(IN^2V^2)$ per wavenumber $k$, which can lead to approximately $N+V$ times faster computations compared to DEISM. As the frequency increases, larger $N$ or $V$ is usually necessary to accurately describe the directivity. Therefore, one might find DEISM-LC more practical for simulating high frequencies.

While the simplified method described here is similar to that of the FSRR method, we note that there are significant differences between Eq.~\eqref{eq:DEISMLCRTF} and Eq.~\eqref{eq:FSRRRTF}. As stated in Sec.~\ref{sec:31}, the FSRR utilizes directivity coefficients $\tilde{C}^{(\text{s})}_{n,m}(k),\tilde{C}^{(\text{r})}_{v,u}(k)$ that are obtained at $1$~m from the source or the receiver. However, the directivity coefficients used in DEISM-LC are not limited to this assumption as long as the transparent spheres enclose the devices. Furthermore, there are inherent dissimilarities between DEISM-LC and FSRR due to the different coefficients, e.g., $B_{\mathbf{p},\mathbf{q}}, M_{\mathbf{p},\mathbf{q}}$ used in Eq.~\eqref{eq:FSRRRTF} and $i^n,i^v$ used in Eq.~\eqref{eq:DEISMLCRTF}.  This difference is also illustrated by an example described in Sec.~\ref{sec:41b}.

\subsection{Algorithm Summary}
\label{sec:35}


A summary of the algorithm proposed in Sec.~\ref{sec:33} is listed as pseudocode in \ref{alg:precals}, \ref{alg:imagegen} and \ref{alg:DEISM_main}. 
A \texttt{Python} implementation of the algorithm, along with an example, is available online at \url{https://github.com/audiolabs/DEISM}. The pseudocode of the far-field approximation method is not provided here for brevity.

\begin{algorithm}[h]
	\caption{Calculate lookup tables}\label{alg:precals}
	\begin{algorithmic}[1]
		\STATE Calculate the Wigner-3j symbols $W_1(n,v,l)$ and $W_2(n,v,l,m,u)$ from Eq.~\eqref{eq:Wigners}.
		\STATE Calculate the spherical harmonic coefficients $P_{n,m}(k, r_0)$ shown in Eq.~\eqref{eq:sampledJpoints} for both the source and receiver using the least squares approach. 
		\STATE Calculate the directivity coefficients of the source and receiver, viz. $C^{(\text{s})}_{n,m}(k)$ and $C^{(\text{r})}_{n,m}(k)$, using the relation $P_{n,m}(r_0,k) = C_{n,m}(k) h_n (k r_0)$. Then store the directivity coefficients as matrices $C^{(s)}(n,m,k)$ and $C^{(r)}(n,m,k)$.
	\end{algorithmic}
\end{algorithm}

The calculation of the Wigner 3j symbols can be computationally expensive if they are computed for each reflection path. Since they are identical for each reflection path, to reduce the computational effort, we precalculate them for all the given spherical harmonic orders and modes $n,m,v,u,l$, and store them in a lookup table. In addition, the directivity coefficients $C^{(\text{s})}_{n,m}(k)$ and $C^{(\text{r})}_{n,m}(k)$  are precalculated and stored in a lookup table. The procedures for generating the lookup tables are listed in Alg.~\ref{alg:precals}.

In the provided implementation, we separate the whole procedure of the ISM into two parts, viz. the image generation given in Alg.~\ref{alg:imagegen} and the single-path transfer function calculation using DEISM given in Alg.~\ref{alg:DEISM_main}. The computation in the spherical harmonic domain is generally also expensive, and therefore we use parallel computation for all reflection paths after all images are calculated.

\begin{algorithm}[h]
\caption{Image generation}\label{alg:imagegen}
\begin{algorithmic}[1]
\STATE $\mathbf{p} = (p_x,p_y,p_z) \in \mathcal{P} = \{0,1\}^3$ \COMMENT{A triple introduced in Tab.~\ref{tab:tabparameters}}.
\STATE $\mathbf{q} = (q_x,q_y,q_z) \in \mathcal{Q} = \{ -N_m, ... , 0,...,N_m\}^3$ \COMMENT{A triple introduced in Tab.~\ref{tab:tabparameters}}.
\STATE $\mathcal{A} = \mathcal{P} \times \mathcal{Q}$ \COMMENT{A set containing all possible images.}
\[ \]
\FOR{$(\mathbf{p},\mathbf{q}) \in \mathcal{A} $}
\IF{$\| 2q_x - p_x\| + \| 2q_y - p_y\| + \| 2q_z - p_z\| \leq N_o$} \COMMENT{$N_o=$ maximum reflection order.}
\STATE Calculate the vector pointing from the source image to the receiver $\mathbf{R}^{\text{sI}\to\text{r}}_{\mathbf{p},\mathbf{q}}$ in Eq.~\eqref{eq:sourceimagetoreceiver} and its spherical coordinates $(d_{x_0},\theta_{x_0},\phi_{x_0})$.
\STATE Calculate the incident angles $\theta_{a}(\mathbf{p},\mathbf{q})$ w.r.t different walls using Eq.~\eqref{eq:inc_angles}.
\STATE Calculate the reflection coefficients $\beta_{x_1}, \beta_{x_2}, \beta_{y_1}, \beta_{y_2}, \beta_{z_1}, \beta_{z_2}$ using Eq.~\eqref{eq:angdeprefcoef}. 
\STATE Calculate the wall attenuation $\beta(\mathbf{p},\mathbf{q}) = \beta_{x1}^{|q_x-p_x|} \beta_{x2}^{|q_x|} \beta_{y1}^{|q_y-p_y|} \beta_{y2}^{|q_y|} \beta_{z1}^{|q_z-p_z|} \beta_{z2}^{|q_z|}$.
\ELSE
\STATE $\mathcal{A} = \mathcal{A} \setminus \{ (\mathbf{p},\mathbf{q})\}$ \COMMENT{Remove current image if it exceeds $N_o$.}
\ENDIF
\ENDFOR
\end{algorithmic}
\end{algorithm}

\begin{algorithm}[!h]
\caption{DEISM core}\label{alg:DEISM_main}
\begin{algorithmic}[1]
\STATE $P(k)=0$ \COMMENT{Initialize the sound field at $\mathbf{x}_{\text{r}}$.}
    \FOR{$(\mathbf{p},\mathbf{q}) \in \mathcal{A} $}
        \FOR{$n=0$ to $N$}
			\FOR{$m=-n$ to $n$}
			\STATE $\Upsilon = (-1)^{(p_y+p_z)m + p_z n}$  \COMMENT{Part of mirror effect of spherical harmonics in Eq.~\eqref{eq:GISMreverbmodecoupling}.} 
			\STATE $m^\prime = (-1)^{p_x +p_y} m$ \COMMENT{The modified mode $m$ of $\alpha_{v,u}^{n,m}$ in Eq.~\eqref{eq:GISMreverbmodecoupling}.} 
				\FOR{$v=0$ to $V$}
				\FOR{$u=-v$ to $v$}
                   \STATE $\alpha^{n,m^\prime}_{v,u}=0$ \COMMENT{Initialize the mode coupling coefficient in Eq.~\eqref{eq:GISMmodecoupling}.}
					\FOR{$l=|n-v|$ to $n+v$}
					\STATE $\xi = \sqrt{(2n +1)(2v+1)(2l+1)/4\pi} $.
					\STATE $\alpha^{n,m^\prime}_{v,u} \mathrel{+}= 4\pi i^{v-n} (-1)^{m^\prime} i^l  h^{(2)}_l(k d_{x_0}) Y_{l,m^\prime-u}(\theta_{x_0},\phi_{x_0}) W_1(n,v,l)$ 
                    \STATE $\times W_2(n,v,l,m^\prime,u) \xi$ \COMMENT{Update the mode coupling coefficient.}
					\ENDFOR
					\STATE $P(k) \mathrel{+}= \Upsilon \beta(\mathbf{p},\mathbf{q}) C^{(s)}(n,m,k) \alpha^{n,m^\prime}_{v,u} \frac{i (-1)^u}{k} C^{(r)}(v,u,k) $\COMMENT{Summation over $\mathcal{A}$ constitutes the room transfer function in Eq.~\eqref{eq:DEISMRTFreverbfinal}.}
				\ENDFOR
				\ENDFOR
			\ENDFOR
			\ENDFOR
		\ENDFOR
\end{algorithmic}
\end{algorithm}

\newpage

\section{Evaluation of the proposed methods}
\label{sec:4}

In the following sections, we describe the experimental setups used to evaluate the performance of DEISM and DEISM-LC (see Eq.~\eqref{eq:DEISMRTFreverbfinal} and Eq.~\eqref{eq:DEISMLCRTF}, respectively). The simulated RTFs are plotted and compared for different scenarios. The differences between the DEISM and the FEM solutions are evaluated using error metrics, and the difference between DEISM and DEISM-LC is analyzed. In addition, the effects of incorporating directivities into the RTFs by comparing the DEISM-LC with the original ISM, and the differences between DEISM-LC and FSRR are also demonstrated by comparing them with FEM.

\subsection{Test setup}
\label{sec:41}

The proposed methods require a description of the directivities of the devices, i.e., a loudspeaker (or microphone) in the spherical harmonic domain. These data can be obtained from measurements of the pressure field in the local vicinity of the transducer and its enclosure. Alternatively, directivity patterns obtained from simulations could be used. In this work, the FEM is used to simulate the free-field pressure fields surrounding the transducer and its enclosure.
Aretz et al.~\cite{aretz2014application} have demonstrated that solutions of the ISM with angle-dependent reflection coefficients are in good agreement with FEM solutions above the Schroeder frequency. Therefore, the FEM is used in this work to evaluate the performance of the proposed method. 

This section presents the details of generating the simulated transducer directivities and the RTFs.

\subsubsection{Transducer directivities}
\label{sec:411}

\begin{figure}[t]
    \figline{
    \fig{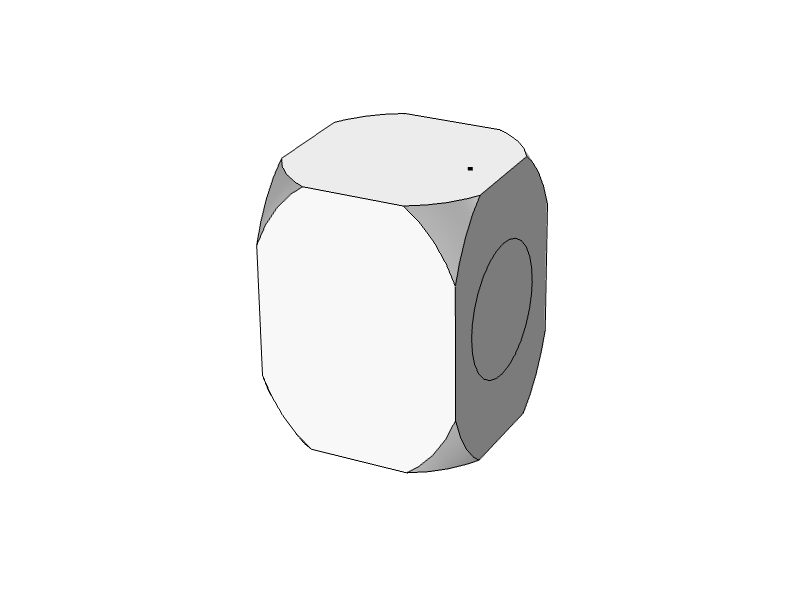}{0.15\textwidth}{}
    \fig{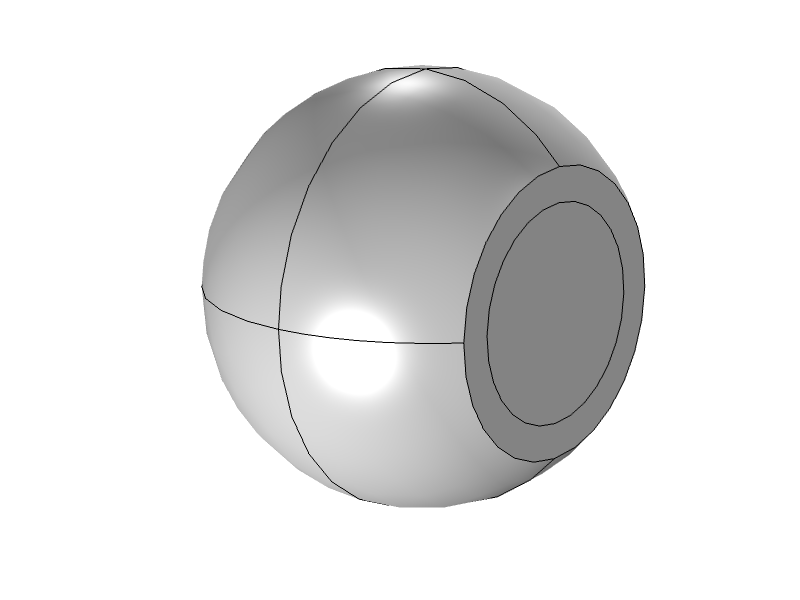}{0.1\textwidth}{}
    \fig{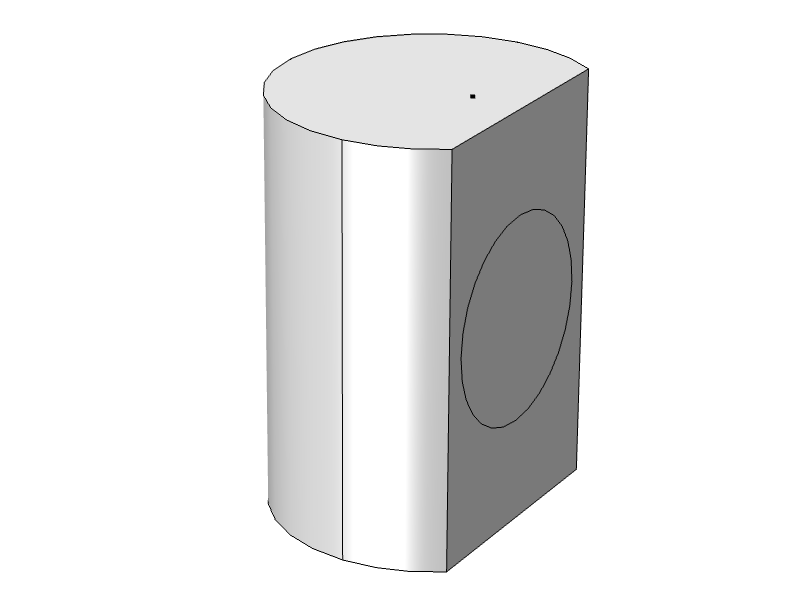}{0.1\textwidth}{}
    }
    \caption{Speaker shapes considered in this work. From left to right, these speakers are referred to in the text as: cuboidal (cub.), spherical (sph.) and cylindrical (cyl.), where the terms in brackets are abbreviations also used in this work. The three speakers share the same height and the same-sized vibrating piston. }
    \label{fig:speaker shapes}
\end{figure}

The three speaker shapes considered in this work are shown in Fig.~\ref{fig:speaker shapes}. For each speaker, the loudspeaker driver is modeled as a vibrating piston with frequency-independent unit acceleration. For each microphone, a monopole source with frequency-independent unit acceleration is placed at the microphone position, which is located on the top and close to the front of each enclosure. 

One must ensure that free-field conditions are used to capture the loudspeaker and microphone directivities. In a measurement setup, one might use an anechoic chamber. When using the FEM, a suitable non-reflecting boundary condition is required.
In this work, a perfectly matched layer~\cite{Berenger_1994} is used to reduce unwanted reflections from the edge of the computational domain. Note that care must be taken to ensure that the perfectly matched layer sufficiently reduces the unwanted reflections - this is achieved in this work by performing \emph{a priori} mesh studies.
Each speaker is meshed using triangular elements, the surrounding air (with a sound speed of 343~m/s and density of 1.2~kg/m$^3$) is meshed using tetrahedral elements, and the perfectly matched layer is meshed using hexahedral elements. An average element size that ensures that there are ten degrees of freedom (or five quadratic elements) per wavelength is used. Quadratic shape functions are used. The free-field transfer function is simulated from 20~Hz to 1 kHz with steps of 2~Hz.

For each speaker, two sizes are considered, which are referred to as Size~1 and Size~2, where Size~2 is the smallest. The sound pressure field is simulated on a transparent sphere around the speaker for each size and shape. 
Note that, for a given speaker shape, the radius of the transparent sphere changes depending on the size of the speaker and the position of the transducer on the speaker; The reason being that we place the transducer at the center of the coordinate system. For the source on the speakers of Size~1, the radius of the transparent sphere is 0.4~m, while for Size~2, it is 0.2~m. For the receivers on the speakers, a radius of 0.5~m is used for Size~1, and a radius of 0.25~m is used for Size~2.

The complex-valued pressure field simulated on the surface of the transparent sphere is used to compute the spherical harmonic directivity coefficients. For the loudspeaker directivity, this involves a straightforward implementation of spherical harmonic decomposition~\cite{Williams1999}. To obtain the receiver directivities, we use acoustic reciprocity~\cite{Xu_et_al_2022} (see Sec.~\ref{sec:32}). 

\begin{figure}[!h]
    \figline{
    \fig{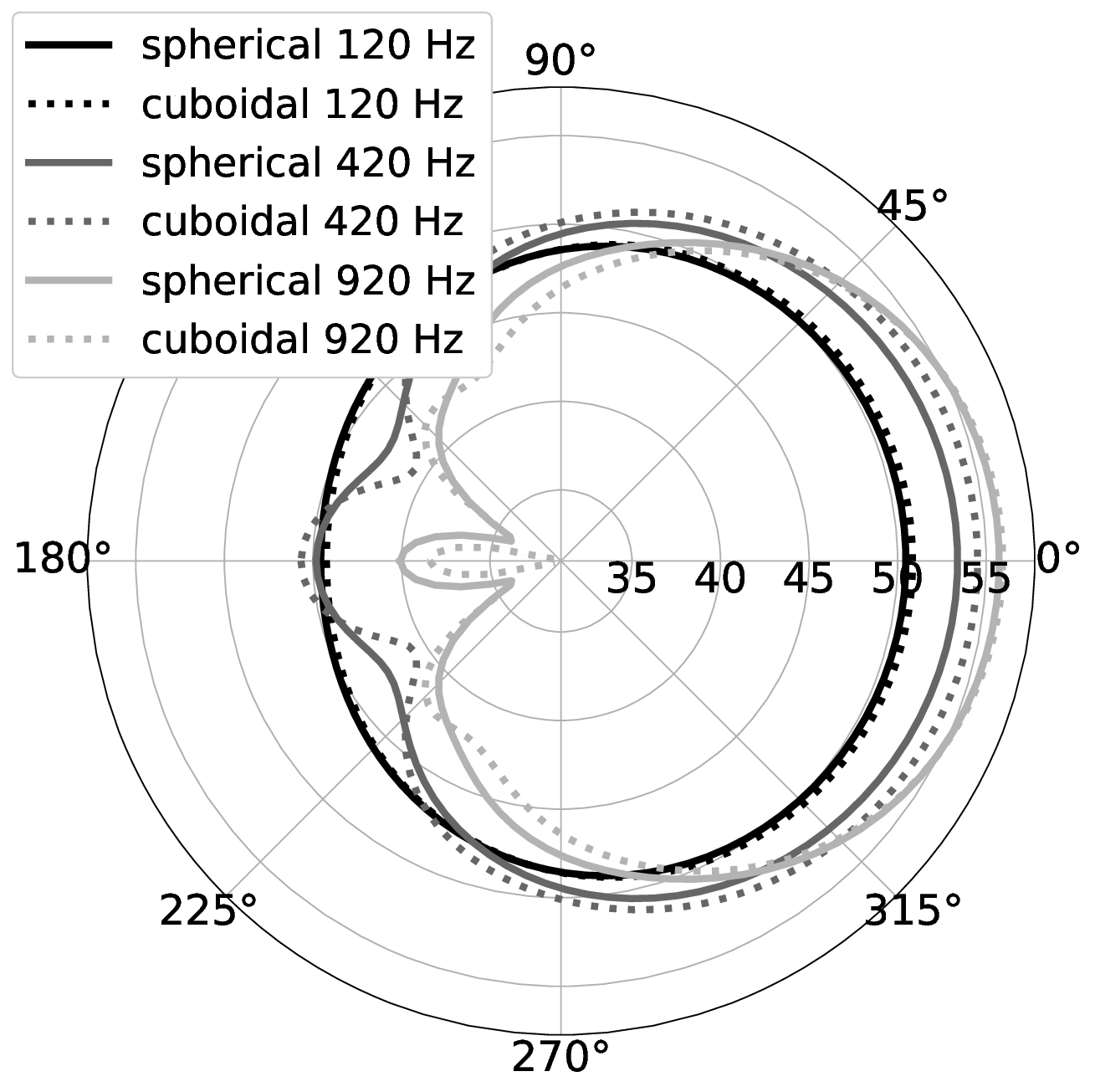}{0.175\textwidth}{Sound pressure level~(dB)}
    \fig{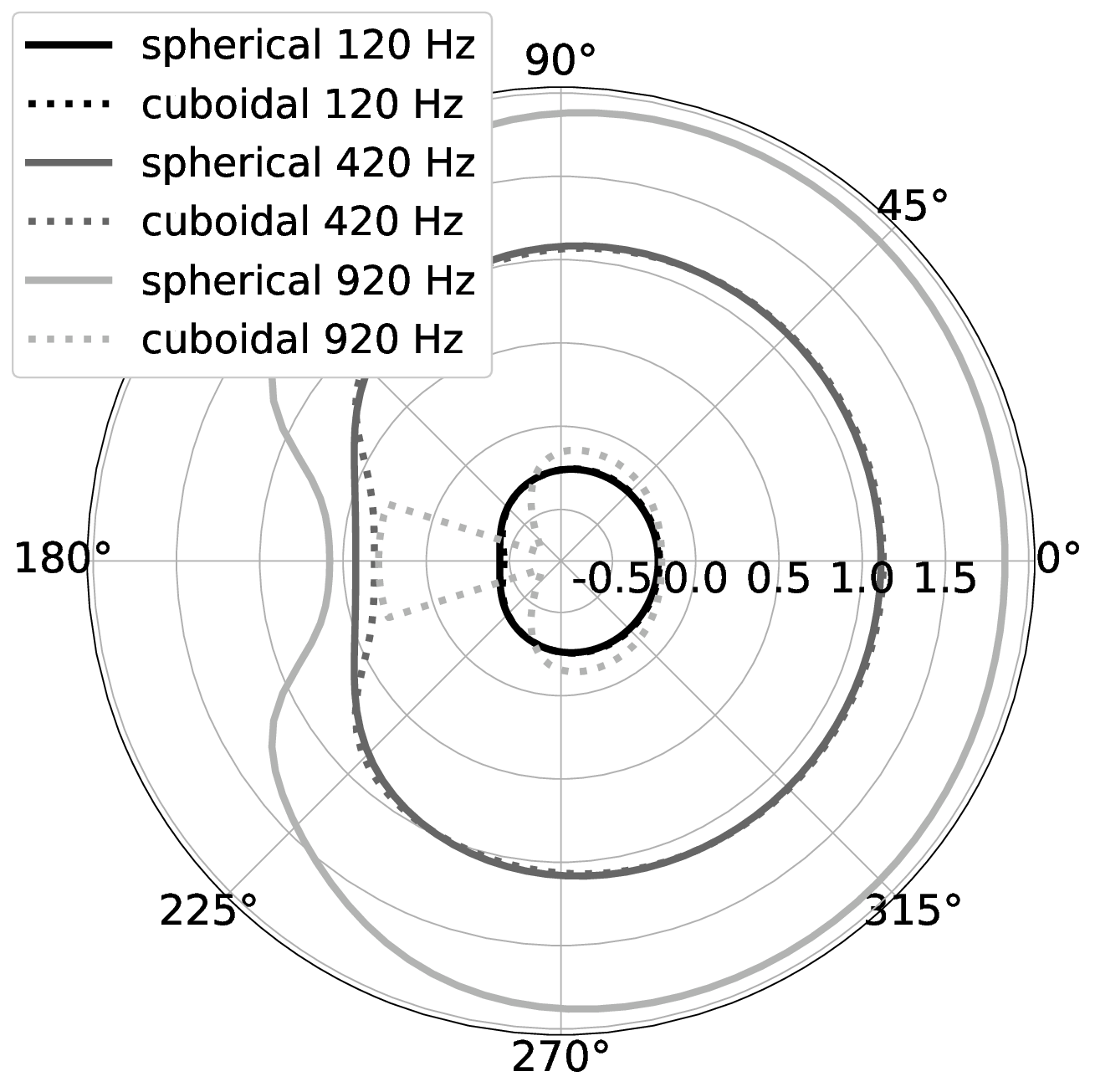}{0.175\textwidth}{Phase in radians~($\pi$)}
    {}
    }
    \caption{Plots of sound pressure level and phase reconstructed using simulated speaker directivity coefficients of a large (Size~1) spherical and cuboidal speaker. The pressure field is reconstructed with maximum spherical harmonic order $5$. The results are evaluated on the XY-plane at three frequencies.}
    \label{fig:DirectivityExample}
\end{figure}

An example of the directivity patterns is shown in Fig.~\ref{fig:DirectivityExample}, where the Sound Pressure Level (SPL) and phase are plotted for the spherical and cuboidal speakers with Size~1. The vibrating piston faces the positive direction of the $x$-axis, which is hereafter referred to as $+x$. The pressure field on the XY-plane is reconstructed using Eq.~\eqref{eq:sampledJpoints} using simulated directivity coefficients $C^{(\text{s})}_{n,m}$ with maximum spherical harmonic order $N=5$. It can be seen that both the SPL and phase plots show differences between the two selected speaker shapes, especially at angles that are in the shadow region of the speakers. One can also observe a significant phase jump in the shadow region when a cuboidal speaker is used, which is less prominent for a spherical speaker. 

\subsubsection{Room transfer functions}
\label{sec:412}

A rectangular room with dimensions $L_x \times L_y \times L_z = 4~\text{m} \times 3~\text{m} \times 2.5~\text{m}$ is considered. The air in the room has a speed of sound of 343~m/s and a density of 1.2~kg/m$^3$. A uniform frequency-independent normalized impedance of 18, which gives an absorption coefficient of approximately 0.2, is specified at the walls of the room. This value is chosen because it gives a reverberation time of approximately 0.29~s at the fundamental frequency of 43.88~Hz, which gives a Schroeder frequency of approximately 198~Hz. 
The reverberation time was estimated using the modal model described by Morse and Bolt~\cite{Morse_and_Bolt_1944}.
Note that, in general, the surface impedance is a complex-valued frequency-dependent quantity. Our choice of a real-valued frequency-independent impedance does not limit the generality of the method described here.

\begin{figure}[!h]
	\centering
	\fig{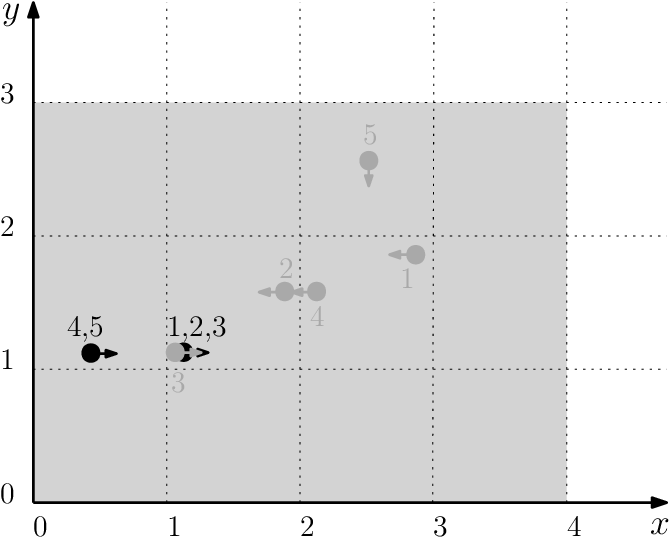}{0.4\textwidth}{}
	\caption{A 2D view of the positions of the mounted transducers inside the room. The black dots and gray dots represent the approximate positions of the source and receiver transducers, respectively. The arrows of the dots indicate the directions in which the speakers face, and the numbers indicate the source and receiver positions.}
	\label{fig:2Droomsetup}
\end{figure}

Five source-receiver configurations are simulated. The center position of each transducer, for each configuration, is given in Table~\ref{tab:speakerpos}. Four of the configurations contain two devices, one with a loudspeaker and the other with a microphone. 
The remaining configuration 3 has a single device on which the loudspeaker and microphone are placed. 
The transfer functions of the direct path between the source and receiver are simulated using the FEM for Configuration 3. 
The microphone position $\mathbf{x}^{\text{C3}}_{\text{r}}$ in Table~\ref{tab:speakerpos} depends on the size, shape, and rotation of the speaker and takes the values given in Table~\ref{tab:xrpos3}. A 2D illustration of the speaker position and facing configurations is shown in Fig.~\ref{fig:2Droomsetup}. The source and receiver are both at least $1$~m away from the walls in Positions $1-3$. Positions $4-5$ allow a closer distance between the walls and at least one speaker to examine the performance of the DEISM in challenging configurations.

\begin{table}[t]
\caption{Configurations of the transducers and the corresponding speaker orientations used in the room simulation. The position vector $\mathbf{x}^{\text{C3}}_{\text{r}}$ for Configuration~3 is listed for different shapes and sizes in Tab.~\ref{tab:xrpos3}.}
\centering
\begin{ruledtabular}
\begin{tabular}{lcc}
Configuration id  & source, orientation & receiver, orientation  \\
\hline
1 & [1.1, 1.1, 1.3], $+x$ & [2.9, 1.9, 1.3], $-x$  \\
2 & [1.1, 1.1, 1.3], $+x$ & [1.9, 1.6, 1.4], $-x$  \\
3 & [1.1, 1.1, 1.3], $+x$ & $\mathbf{x}^{\text{C3}}_{\text{r}}$, $+x$  \\
4 & [0.4, 1.1, 1.3], $+x$ & [2.1, 1.6, 1.3], $-x$  \\
5 & [0.4, 1.1, 1.3], $+x$ & [2.5, 2.6, 1.3], $-y$  \\
\end{tabular}
\end{ruledtabular}
\label{tab:speakerpos}
\end{table}

\begin{table}[t]
\caption{Positions $\mathbf{x}^{\text{C3}}_{\text{r}}$ for simulations with the source and receiver on the same speaker (Configuration~3).}
\centering
\begin{ruledtabular}
\begin{tabular}{lcc}
Speaker shape  & Size~1 (large) & Size~2 (small)  \\
\hline
Cuboid & [1.05, 1.1, 1.5] &  [1.075, 1.1, 1.4] \\
Spherical & $[1.05, 1.1, 1.473]$ & $[1.075, 1.1, 1.387]$  \\
Cylindrical & [1.05, 1.1, 1.5] & [1.075, 1.1, 1.4]  \\
\end{tabular}
\end{ruledtabular}
\label{tab:xrpos3}
\end{table}

Each model is meshed using tetrahedral elements with an average size that ensures that 10 degrees of freedom per wavelength are used. Quadratic shape functions are used. The room transfer functions for each configuration are simulated, from 20~Hz to 1 kHz, in steps of 2~Hz.

\subsection{Preliminary comparisons with ISM and FSRR}
\label{sec:41b}
\begin{figure}[t]
     \centering \figcolumn{
    \fig{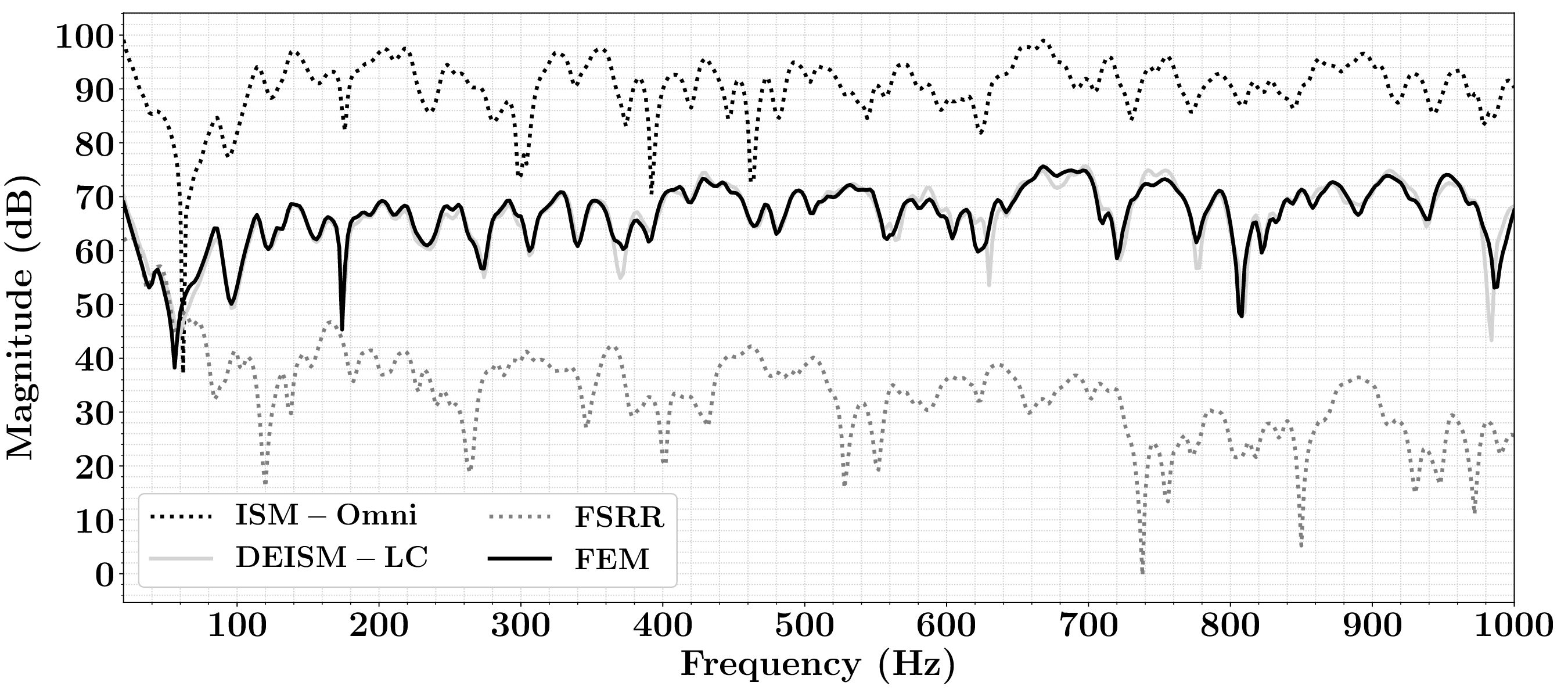}{0.5\textwidth}{}
    \fig{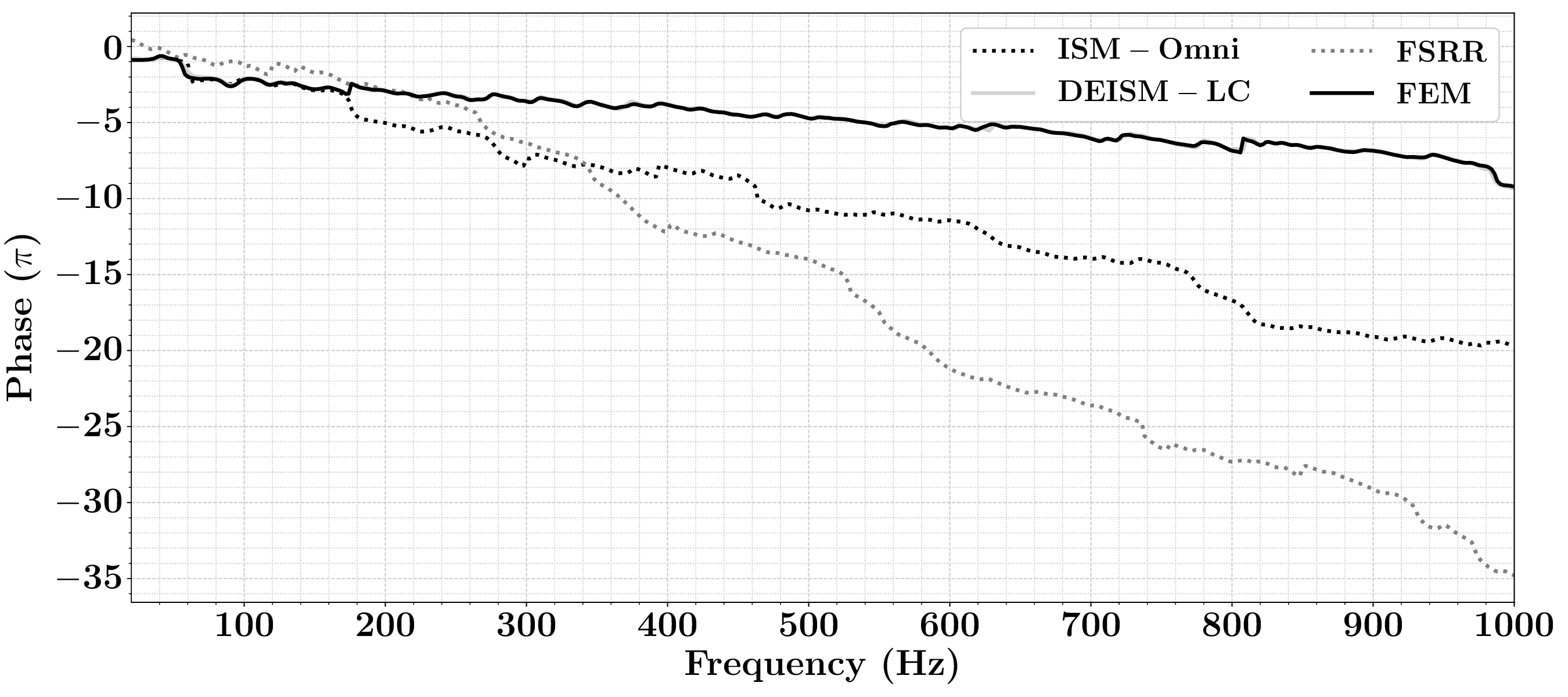}{0.5\textwidth}{}
     }
    \caption{A comparison of the magnitudes and phase responses between the ISM-Omni, DEISM-LC, FSRR, and FEM solutions. The results are shown for large cuboidal speakers with Configuration 2. Max. reflection order is 25 and max. SH orders are 5 for both the source and receiver.}
    \label{fig:DEISMLCFSRRISM}
\end{figure}

In this section, we compare the DEISM-LC to existing RTF simulation methods. To this end, we choose the original ISM with omnidirectional transducers (referred to as ISM-Omni) and the FSRR method, as introduced in Eq.\eqref{eq:ism} and Eq.\eqref{eq:FSRRRTF}, respectively. We demonstrate the effects of incorporating transducer directivities into the RTFs by comparing the ISM-Omni with the DEISM-LC. We are also interested in the differences between the FSRR method and DEISM-LC as they are computationally more efficient compared to the DEISM.
In Fig.~\ref{fig:DEISMLCFSRRISM}, the magnitudes and phases of the RTFs using the ISM-Omni, DEISM-LC, FSRR, and FEM are plotted. As an example, the results are shown for large cuboidal speakers in Configuration 2 with maximum reflection order 25 and maximum spherical harmonic order 5 for both the source and receiver. The magnitudes of the RTFs are given in terms of SPL in decibels and the unwrapped phases in radians. Unless specified, the magnitude and phase responses use the same units in the subsequent sections.
The frequency-dependent SPL is defined as \cite{Kuttruff_2017} 
\begin{equation}
    \text{SPL}(k) = 20 \log_{10} \bigg(\frac{P_{\text{rms}}(k)}{P_0} \bigg)~,
\end{equation}
where $P_{\text{rms}}$ is the root-mean-square pressure calculated from the RTFs and $P_0 = 2 \times 10^{-5}~\text{Pa}$ is the reference pressure. The FEM solutions are presumed as the ground truth of the RTFs in this section and throughout the following evaluations. 

The RTFs of the ISM-Omni were generated using an omnidirectional source and receiver at the same locations as those used in Configuration 2. The directivity coefficients of the source and receiver are the same and contain only a monopole component based on Eq.~\eqref{eq:monopolereceiverC}, i.e., 
\begin{equation}
    C^{(\text{s})}_{0,0}(k) = C^{(\text{r})}_{0,0}(k) = - i k j_0(0) Y_{0,0}^*(\theta,\phi) = \frac{-ik}{\sqrt{4\pi}}~,
\end{equation}
where $(\theta,\phi)$ denotes the directions in spherical coordinates of either the point source or the omnidirectional receiver relative to their local origins. For the RTFs of the FSRR given in Eq.\eqref{eq:FSRRRTF}, the directivity coefficients $\tilde{C}^{(\text{s})}_{n,m}(k),\tilde{C}^{(\text{r})}_{n,m}(k)$ are extrapolated from $C^{(\text{s})}_{n,m}(k),C^{(\text{r})}_{v,u}(k)$ to the measurement sphere with a radius of $1$~m, and we choose $M_{\mathbf{p},\mathbf{q}} = \beta(\mathbf{p},\mathbf{q})$. 

Overall, the differences between the DEISM-LC and FEM are small compared to those between the ISM-Omni and FEM and those between the FSRR and FEM. This indicates that the DEISM-LC can generate more realistic RTFs when compared to the other image source methods for the considered scenario. The large offset in magnitude responses between the DEISM-LC and ISM-Omni mainly come from the different source types, i.e., a point source used in ISM-Omni and a vibrating piston used in DEISM-LC. However, the locations and Q-factors of the resonant peaks and the phase responses also vary greatly. This shows that the directivities of the source and receiver, including sound diffraction around the speakers, can affect the RTFs to a large extent. 

The differences between the DEISM-LC and FSRR are mainly attributed to the inherent mismatch in their RTF parameterizations, i.e., between Eq.\eqref{eq:DEISMLCRTF} and Eq.\eqref{eq:FSRRRTF}. 
The magnitude responses are different not only in the amplitude but also in the locations and Q-factors of the resonant peaks. The phase response of the FSRR even contains positive values at lower frequencies corresponding to non-causal behaviors, which results from the random sign $B_{\mathbf{p},\mathbf{q}}$ used in Eq.~\eqref{eq:FSRRRTF}. 

Since the inherent differences between DEISM-LC and other methods, i.e., the ISM-Omni and FSRR, are shown to be large, we limit the following evaluations to DEISM, DEISM-LC, and FEM.

\subsection{Varying distance between source and receiver}
\label{sec:42}

\begin{figure}[t]
     \centering\figcolumn{\fig{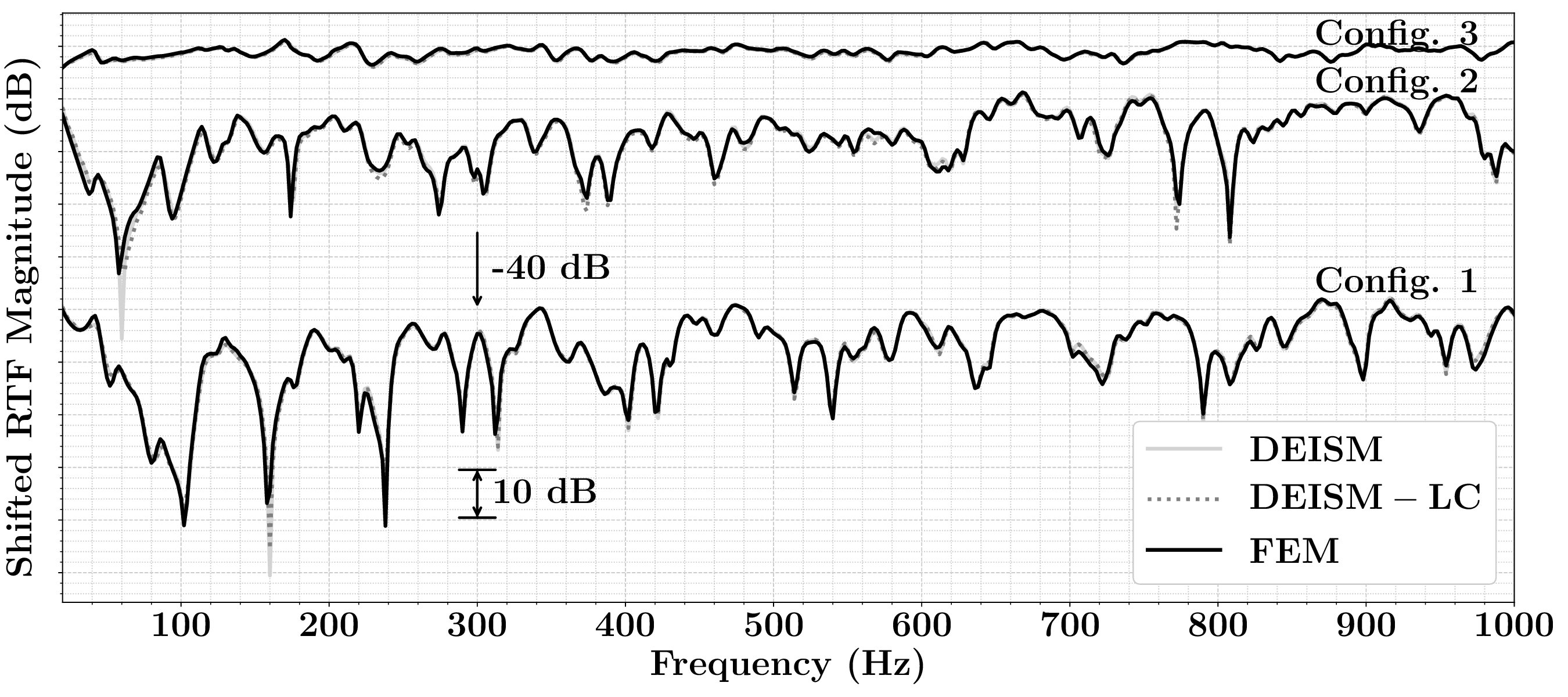}{0.5\textwidth}{}\fig{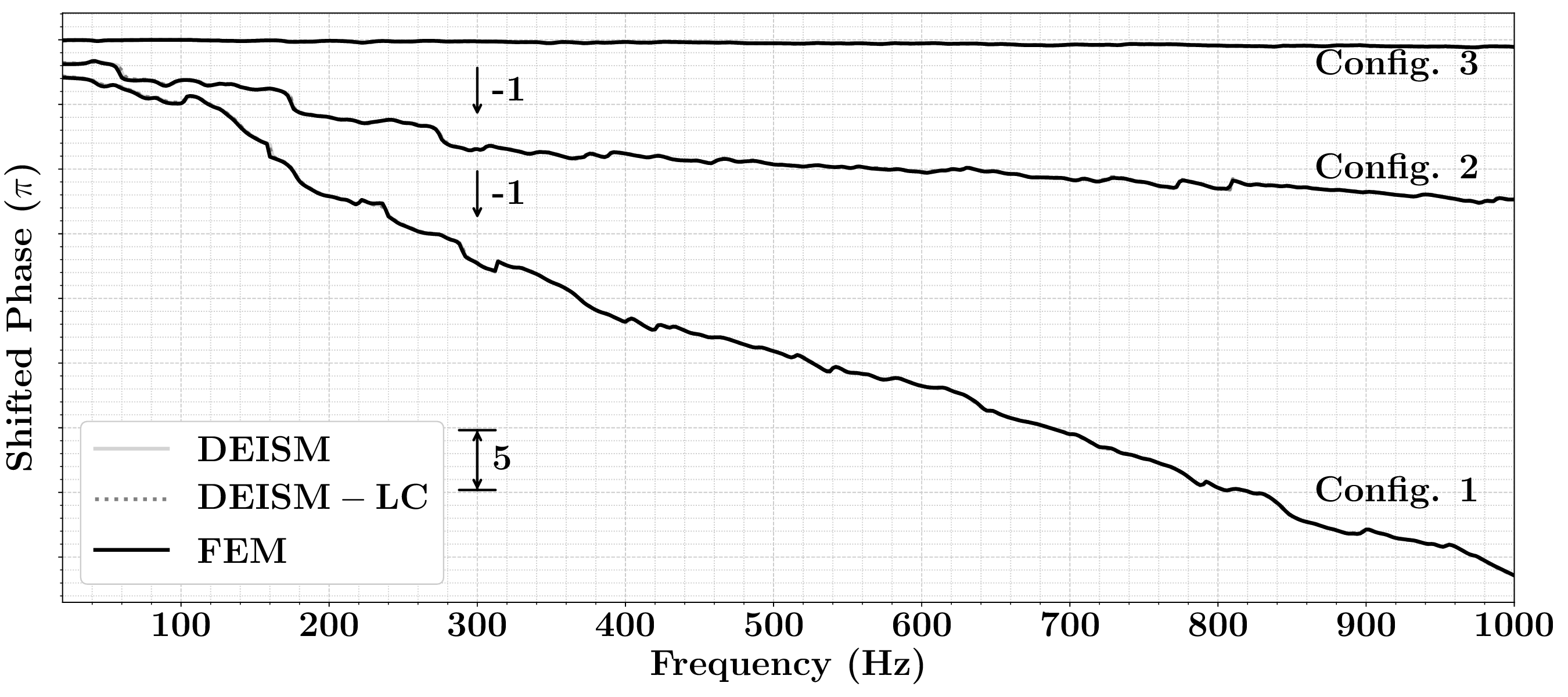}{0.5\textwidth}{}
     }
    \caption{Magnitudes and phase responses for small spherical speakers with varying distance between speakers. Max. reflection order is 25 and max. SH orders are 5 for both the source and receiver. The magnitudes and phases are shifted by vertical arrows for better distinction of the curves. 
    }
    \label{fig:VerifyDEISM1}
\end{figure}

In this section, we present a comparison between the FEM and DEISM solutions for the small spherical speaker when the distance between the source and receiver is changing, corresponding to Configuration 1-3 in Tab.~\ref{tab:speakerpos}. The two distances in Configuration 1-2 are around $1.97$~m, $0.95$~m and the distances in Configuration 3 are around $0.1$~m and $0.2$~m for the small and large speakers, respectively.
The magnitude and phase are compared in Fig.~\ref{fig:VerifyDEISM1}.
To avoid visual overlap of the curves for different configurations, the SPLs and phases are plotted by shifting the curves vertically, as denoted by the arrows.

We see that good agreement is obtained between the FEM and DEISM solutions.
In general, greater differences are found at the notches of the RTFs, than at the peaks.
Additionally, the differences between DEISM and DEISM-LC solutions are much smaller than those between the FEM and DEISM solutions in the notches of the RTFs. Since Configuration~3 contains only one speaker in the room, the corresponding RTFs show fewer variations across frequencies compared to the other two configurations. 
Similar agreement between FEM and DEISM was found for the other speaker shapes. For brevity, the respective plots are not presented here.

\subsection{Source and receiver on the same device}
\label{sec:43}

Next, we compare solutions for cases where the source and receiver are placed on the same device (Configuration~3).
For this comparison, we analyze data for the three speaker shapes considered.

\begin{figure}[t]
     \centering \figcolumn{\fig{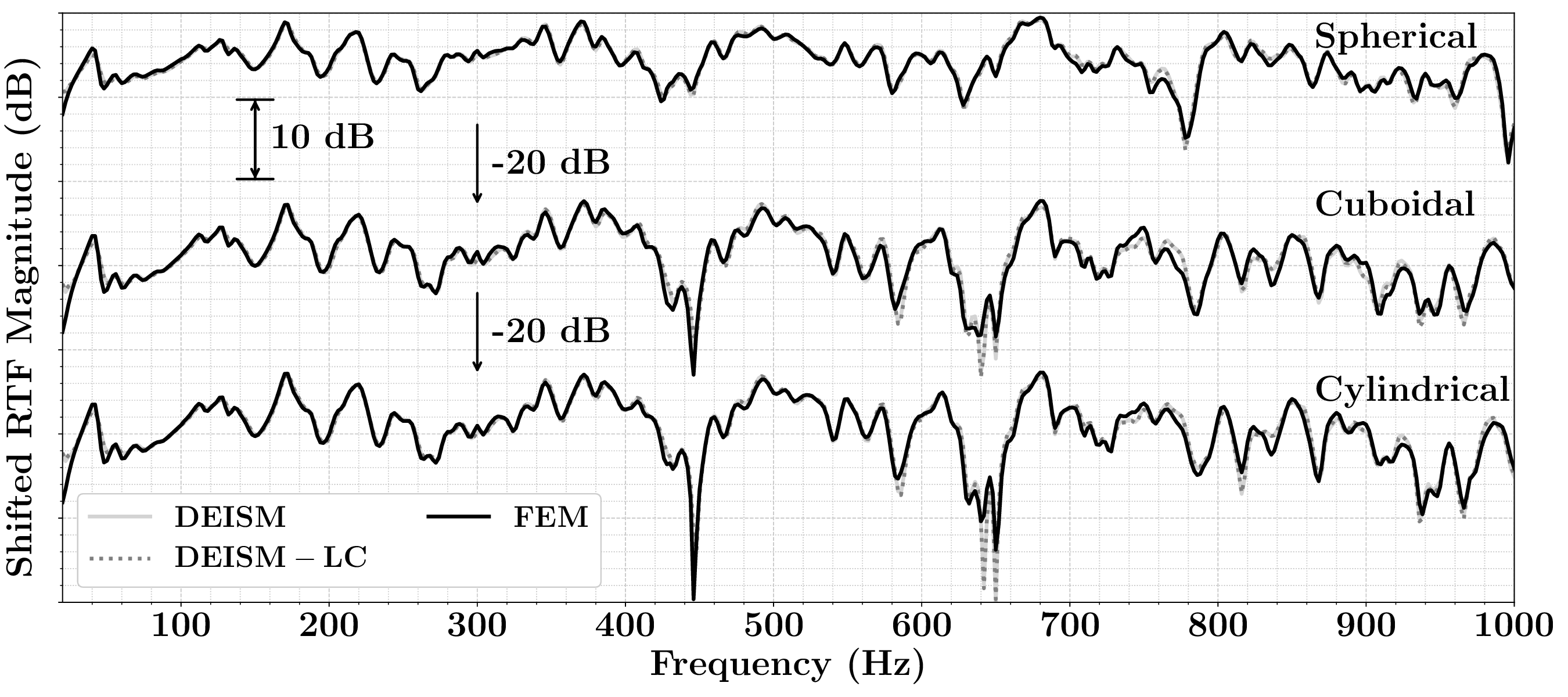}{0.5\textwidth}{}\fig{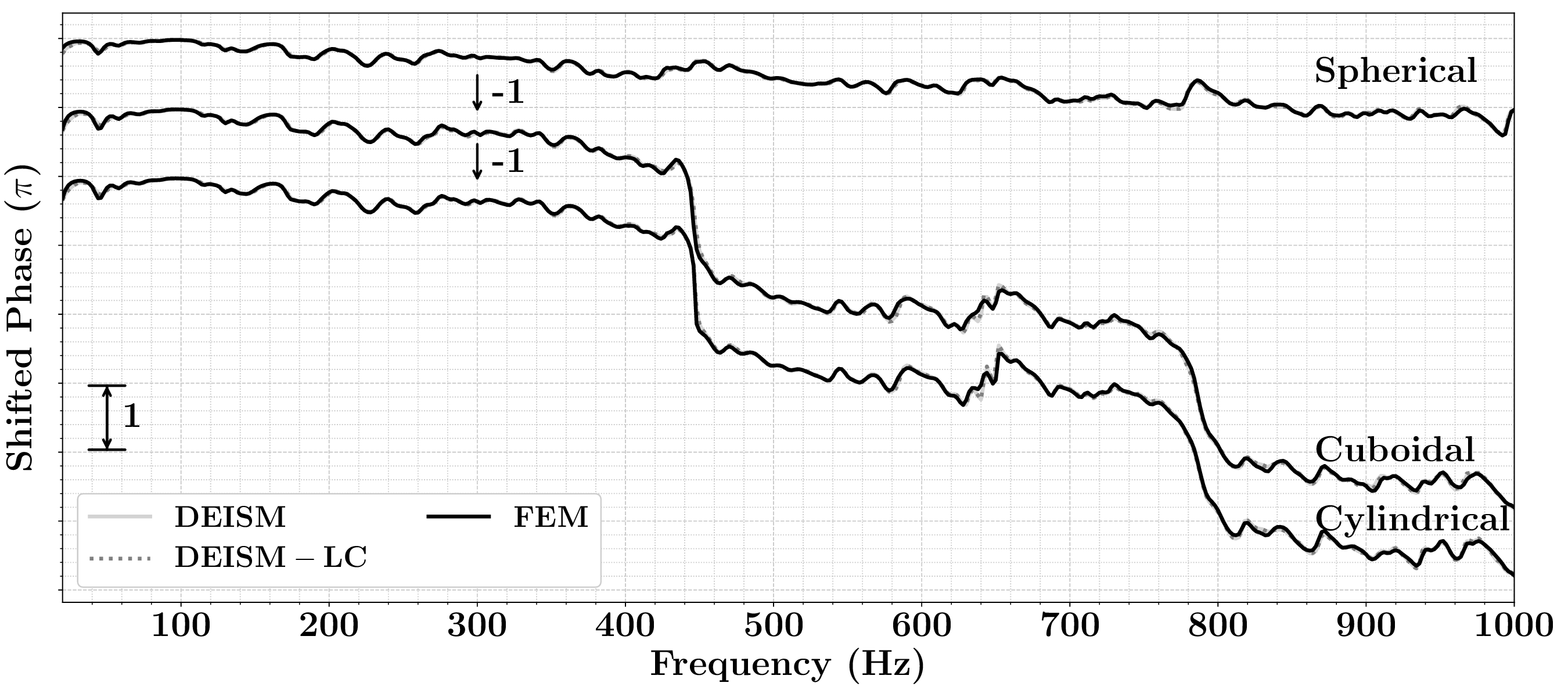}{0.5\textwidth}{}
     }
    \caption{Magnitudes and phase responses for large speakers with Configuration~3 while varying shapes of the speakers. Max. reflection order is 25 and max. SH. orders are 5 for both the source and receiver. The magnitudes and phases are shifted by verticle arrows for better distinction of the curves.}
    \label{fig:VerifyDEISM2}
\end{figure}

The respective magnitude and phase solutions are shown in Fig.~\ref{fig:VerifyDEISM2}. 
Once again, good agreement is obtained between the FEM and DEISM solutions.
The magnitudes and phases of the RTFs of the cuboidal and cylindrical speakers show very similar behaviour. In contrast, those of the spherical speaker exhibit clear differences when compared to the RTFs of the other speakers.  
Note that the main differences appear at the notches of the RTF magnitudes, which correspond to frequencies around $444$~Hz and $650$~Hz. In addition, the phase curves of the cuboidal and cylindrical speakers also exhibit significant jumps at some frequencies, which can not be observed for the spherical speaker. The differences between the curves indicate that the edges of the cuboidal and cylindrical speakers can induce additional diffraction and lead to distinct characteristics in the RTFs.

\subsection{Quantitative analysis}
\label{sec:44}

In this section, we quantify the differences between the FEM and DEISM/DEISM-LC solutions and DEISM and DEISM-LC solutions. The following error metrics are used for the comparison:
The root-mean-square log spectral distance~\cite{Gray1976}
\begin{equation}
e_{\text{lsd}} =\sqrt{ \frac{1}{K} \displaystyle \sum_{k \in \mathcal{K}} \Bigg| 10 \log_{10} \bigg(  \frac{|H_{\text{Test}}(k)|}{|H_{\text{FEM}}(k)|} \bigg)^2 \Bigg|^2 }~,
\end{equation}
and the normalized root-mean-square phase error in radians \cite{Jarrett2012b}
\begin{equation}
e_{\text{p}} = \sqrt{\frac{1}{K} \displaystyle \sum_{k \in \mathcal{K}} \big(\angle H_{\text{Test}}(k)  - \angle H_{\text{FEM}}(k) \big)^2 }~,
\end{equation}
where $\mathcal{K}$ is the set of the $K$ wavenumbers used in simulations, $H_{\text{Test}}(k)$ is the simulated RTF of either the DEISM or DEISM-LC at wavenumber $k$, and $H_{\text{FEM}}(k)$ is the simulated RTF of the FEM at wavenumber $k$.
The log spectral distance and the phase errors are presented for different positions, speaker shapes, and sizes in Tables~\ref{tab:error} and \ref{tab:error_phase}, respectively. In both tables, lower values indicate better performance. General observations can be made for Positions 1-3 in the two tables:
\begin{itemize}
    \item The smaller the speaker is, the smaller the difference between the DEISM and FEM solutions. 
    \item The fewer speakers there are, i.e., for Configuration~3, the smaller the difference. 
    
\end{itemize}

\begin{table}[t]
\caption{Root-mean-square log spectral distance in decibel between DEISM and FEM, and between DEISM-LC and FEM. (s. denotes small.)}
\centering
\begin{ruledtabular}
\begin{tabular}{lcccccc}
config. id & cub. & s. cub. & sph. & s. sph. & cyl. & s. cyl.  \\
\hline
1 DEISM & 1.918 & 1.214 & 2.268 & 1.143 & 2.062 & 0.950 \\
1 DEISM-LC & 1.669 & 1.070 & 2.206 & 1.125 & 2.007 & 0.916 \\
\hline
2 DEISM & 1.974 & 1.051 & 2.521 & 1.003 & 1.900 & 0.964 \\
2 DEISM-LC & 1.855 & 1.003 & 2.060 & 0.984 & 1.809 & 0.943 \\
\hline
3 DEISM & 0.423 & 0.204 & 0.841 & 0.170 & 0.920 & 0.154 \\
3 DEISM-LC & 0.427 & 0.206 & 0.883 & 0.174 & 0.952 & 0.155 \\
\hline
4 DEISM & 7.018 & 8.187 & 6.457 & 7.418 & 6.350 & 7.287 \\
4 DEISM-LC & 6.864 & 8.149 & 6.461 & 7.298 & 6.240 & 7.211 \\
\hline
5 DEISM & 6.887 & 6.810 & 7.077 & 6.309 & 7.011 & 6.425 \\
5 DEISM-LC & 6.792 & 6.807 & 6.986 & 6.277 & 6.968 & 6.397 \\
\end{tabular}
\end{ruledtabular}
\label{tab:error}
\end{table}

\begin{table}[t]
\caption{Normalized root-mean-square phase error in radians between DEISM and FEM, and between DEISM-LC and FEM. (s. denotes small.)}
\centering
\begin{ruledtabular}
\begin{tabular}{lcccccc}
config. id & cub. & s. cub. & sph. & s. sph. & cyl. & s. cyl.  \\
\hline
1 DEISM & 0.258 & 0.179 & 0.307 & 0.174 & 0.278 & 0.162 \\
1 DEISM-LC & 0.265 & 0.153 & 0.300 & 0.176 & 0.267 & 0.181 \\
\hline
2 DEISM & 0.251 & 0.138 & 0.246 & 0.128 & 0.255 & 0.123 \\
2 DEISM-LC & 0.246 & 0.149 & 0.259 & 0.142 & 0.260 & 0.144 \\
\hline
3 DEISM & 0.051 & 0.026 & 0.104 & 0.032 & 0.093 & 0.024 \\
3 DEISM-LC & 0.052 & 0.026 & 0.109 & 0.032 & 0.097 & 0.024 \\
\hline
4 DEISM & 1.796 & 1.764 & 1.834 & 1.776 & 1.801 & 1.778 \\
4 DEISM-LC & 1.791 & 1.763 & 1.828 & 1.774 & 1.781 & 1.776 \\
\hline
5 DEISM & 1.795 & 1.926 & 1.818 & 1.906 & 1.822 & 1.921 \\
5 DEISM-LC & 1.791 & 1.928 & 1.816 & 1.907 & 1.821 & 1.923 
\end{tabular}
\end{ruledtabular}
\label{tab:error_phase}
\end{table}

The errors of the DEISM solutions increase when the speakers are located close to one wall of the room, i.e., for Configurations 4 and 5. The large errors may be attributed to the mirror source approximation since the speakers can have more complicated wave interaction with the nearby walls, which is not considered in this model, and can lead to additional RTF fluctuations. For devices with a physical extent, the assumption of a source image might not hold and could require more complex modeling.

In Configuration~3, the DEISM outperforms DEISM-LC for all speaker shapes and sizes. However, this does not necessarily hold for Configurations~1 and 2. This is mainly caused by the notches of the RTF magnitudes, where the DEISM produces much lower values than DEISM-LC, as shown in Fig.~\ref{fig:VerifyDEISM1} and Fig.~\ref{fig:VerifyDEISM2}.

To provide insight into the performance of the methods in larger rooms, we consider the case in which the room's length in the $x$-dimension is doubled. For this test, only the cylindrical driver is considered in Configurations 1-3. The error levels obtained are presented in Table~\ref{tab:errorroom2}. The error levels for the larger room are similar to the values given in Tables~\ref{tab:error} and \ref{tab:error_phase}, and imply similar conclusions.

\begin{table}[t]
\caption{Root-mean-square log spectral distance in decibel and normalized root-mean-square phase error in radians between DEISM and FEM, and between DEISM-LC and FEM in a larger room. (s. denotes small.)}
\centering
\begin{ruledtabular}
\begin{tabular}{lcccc}
      & \multicolumn{2}{c}{RMS LSD (decibel)} & \multicolumn{2}{c}{NRMS phase error (radians)} \\
\cmidrule(lr){2-3} 
\cmidrule(lr){4-5} 
config. id & cyl. & s. cyl.  & cyl. & s. cyl. \\
\hline
1 DEISM & 1.177 & 0.967 & 0.185 & 0.128 \\
1 DEISM-LC & 1.190 & 1.036 & 0.227 & 0.126 \\
\hline
2 DEISM & 0.987 & 0.928 & 0.123 & 0.131 \\
2 DEISM-LC & 1.188 & 0.972 & 0.153 & 0.119 \\
\hline
3 DEISM & 0.783 & 0.319 & 0.151 & 0.038 \\
3 DEISM-LC & 0.893 & 0.320 & 0.155 & 0.038 \\
\hline
\end{tabular}
\end{ruledtabular}
\label{tab:errorroom2}
\end{table}

\subsection{Evaluation of DEISM-LC}
\label{sec:45}

\begin{figure}[t]
     \centering \figcolumn{
    \fig{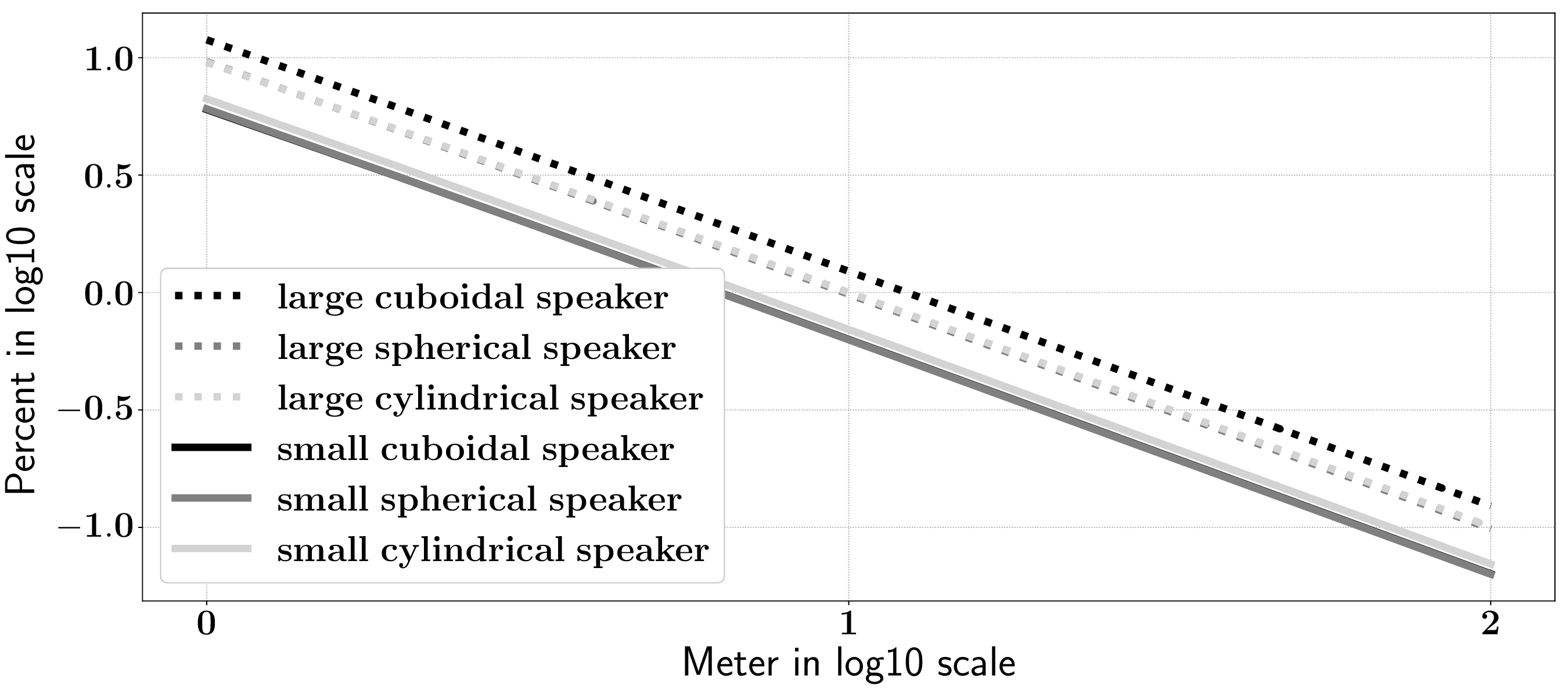}{0.5\textwidth}{}
     }
    \caption{Relative error (Eq.~\eqref{eq:rel_err_l2}) between the DEISM and DEISM-LC solutions evaluated for the direct path only while changing the distance between the source and receiver. Both the distance and relative errors are plotted on a logarithmic scale.}
    \label{fig:simpDEISM1}
\end{figure}

\begin{figure}[t]
     \centering \figcolumn{
    \fig{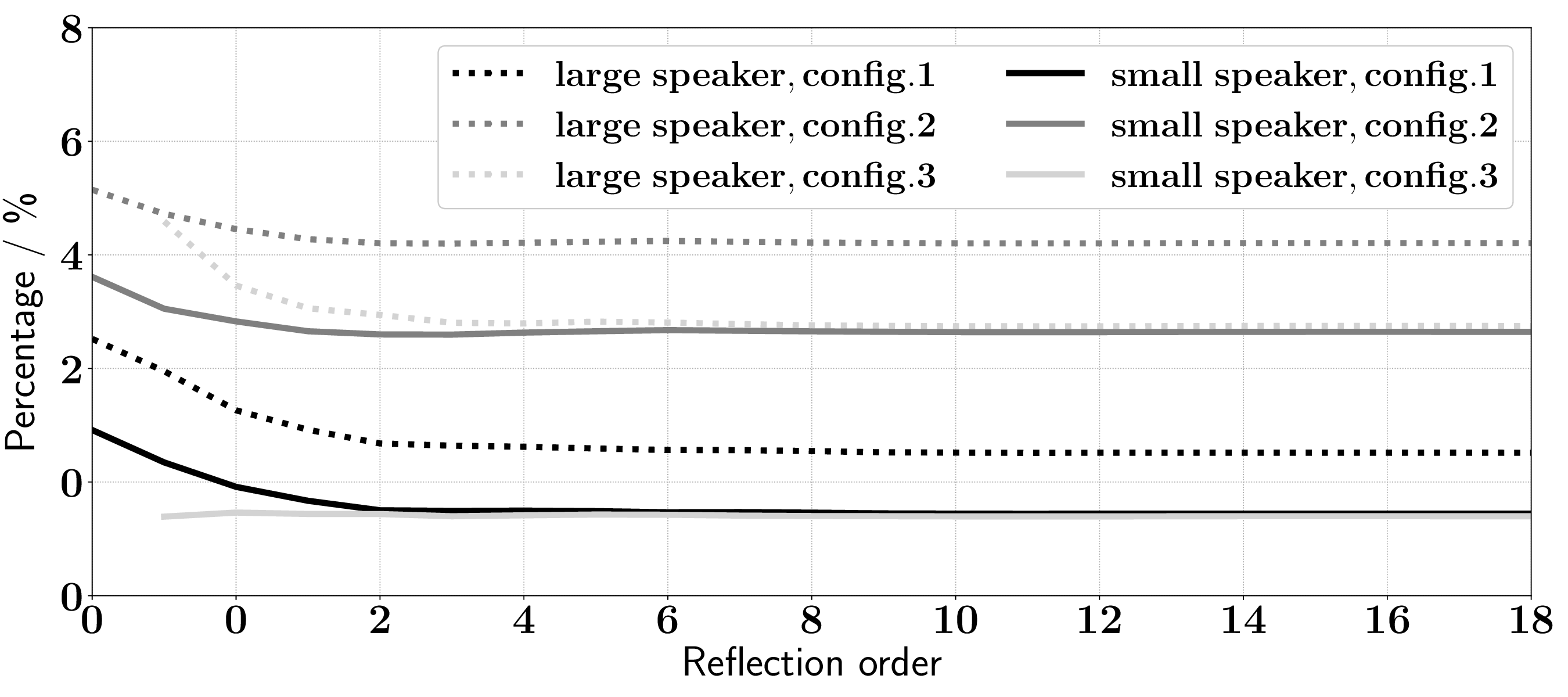}{0.5\textwidth}{}
     }
    \caption{Relative error (Eq.~\eqref{eq:rel_err_l2}) between the DEISM and DEISM-LC solutions as a function of reflection order for the cuboidal speakers at Positions 1, 2, and 3.}
    \label{fig:simpDEISM2}
\end{figure}

In the previous section, we showed that in some cases, DEISM-LC achieves a smaller difference to the FEM than DEISM, especially at the notches of the RTFs. It is worth noting that the frequencies of the notches might not be accurate since the maximum resolution of the frequencies is $2$~Hz. Therefore, directly comparing the DEISMs with the FEM might not be enough to determine whether the simplified DEISM performs worse or better. In the following, we investigate the difference between the DEISMs by computing the relative error using the $\ell ^{2}$ norm, defined as
\begin{equation}\label{eq:rel_err_l2}
e_{\ell^2} = \frac{\|\mathbf{H}_\text{DEISM} - \mathbf{H}_\text{DEISM-LC}\|_2}{\|\mathbf{H}_\text{DEISM}\|_2}~,
\end{equation}
where $\mathbf{H}_{(\cdot)}$ is the complex-valued vector with the transfer function values $H_{(\cdot)}(k)$ of all wavenumbers.
In this comparison, we consider the error in free-field and reverberant conditions.

The errors in free-field conditions are given in Fig.~\ref{fig:simpDEISM1}, where the source-receiver distance and the error are plotted on logarithmic scales. One can observe that the difference decreases to around $1\%$ at a distance of $10$~m and to around $0.1\%$ at a distance of $100$~m. It can be concluded that DEISM-LC asymptotically converges to the DEISM solution as the distance between the source and receiver increases, which is expected from the derivations in Sec.~\ref{sec:34}. 

In Fig.~\ref{fig:simpDEISM2}, the relative error is evaluated by changing the maximum reflection order in the DEISM for the cuboidal speakers. Note that for the curves of Configuration~3, the data of reflection order zero is missing since the direct path is simulated using the FEM. The difference between the DEISM solutions reduces for the first few reflection orders, but remains constant for higher reflection orders. Similar behaviors were also found for other speaker shapes, which are not presented here for brevity. 

\subsection{Summary}
\label{sec:46}

The presented comparisons and error analyses show that the proposed method and its simplified version can simulate RTFs that are in good agreement with FEM solutions in Configurations~1-3. 
Less agreement is found when a speaker is placed close to the walls, i.e., at Configurations~4 and 5. This might be attributed to the comb filtering caused by the direct path and reflection between the wall and speaker when the distance is less than $0.5$~m~\cite{Paasonen2017}.

The errors between the DEISM and FEM solutions decrease when smaller speakers are used in the simulation. Additionally, the lowest error is obtained when the source and receiver are placed on the same speaker, which reduces the number of speakers in the simulation, and hence the number of reflecting surfaces. These observations support the hypothesis that the errors come mostly from the omission of speakers with physical extent in the ISM model, i.e., scattering effects between the walls and the speaker's surfaces are not included. 

We have also shown that the simplified version (DEISM-LC) converges to its complete version (DEISM) in the free field and yields a small bias when used to simulate RTFs. The simplified version, DEISM-LC, yields a better agreement with the FEM simulations as opposed to the existing FSRR method. 

\section{Conclusion}
\label{sec:5}

We have developed a room transfer function simulation method using spherical harmonic directivity coefficients based on the image source method. The directivity coefficients contain information about the sound diffraction around transducers mounted on one or two devices. The proposed method (DEISM) and its simplified version (DEISM-LC) were compared with FEM solutions and evaluated for different acoustic scenarios. The results show that the proposed approaches can simulate room transfer functions with good agreement to FEM simulations when they are not placed close to the room boundaries. A possible rule of thumb to indicate the region inside the room where the DEISM provides results different from FEM is beyond the scope of this paper, and is left as a topic for future studies. 
The shapes and sizes of the speakers can also introduce different acoustic diffraction effects, resulting in additional differences in errors of the room transfer functions. The simplified approach can significantly reduce computational complexity, especially when the frequency increases.

Future work could include extending the proposed method to incorporate the sound reflections between the device enclosures or between the enclosures and room boundaries. Such extensions might reduce the error incurred when devices are placed close to the room boundaries or when multiple devices are used. \tc{Another valuable extension of the proposed approach would be to apply the DEISM and DEISM-LC to rooms with more general geometries, with the aim of efficiently generating more realistic room acoustics simulations.} 

\bibliography{refs}

\end{document}